\documentclass[
 reprint,
 superscriptaddress,
 preprintnumbers,
 nofootinbib,
 amsmath, amssymb,
 aps,
 prd,
 floatfix,
]{revtex4-2}

\usepackage{aas_macros}
\usepackage{xcolor}
\usepackage{graphicx}

\definecolor{plasmablue}{rgb}{0.050383, 0.029803, 0.527975}

\usepackage{hyperref}
\hypersetup{
    colorlinks=true,
    linkcolor=plasmablue,
    urlcolor=plasmablue,
    citecolor=plasmablue,
    pdftitle={SPT Clusters with DES and HST Weak Lensing. I. Cluster Lensing and Bayesian Population Modeling of Multi-Wavelength Cluster Datasets},
    pdfauthor={Bocquet et al.},
    }

\makeatletter
\def\l@subsection#1#2{}
\def\l@subsubsection#1#2{}
\def\Dated@name{}
\makeatother

\renewcommand{\vec}[1]{\mbox{\boldmath$#1$}}

\newcommand{\dif}{\mathrm{d}}

\newcommand{\invSigmac}{\ensuremath{{\Sigma_\mathrm{crit}^{-1}}}}
\newcommand{\sigmawl}{\ensuremath{{\sigma_{\ln M_\mathrm{WL}}}}}
\newcommand{\Mwl}{\ensuremath{M_\mathrm{WL}}}
\newcommand{\Mhalo}{\ensuremath{M_\mathrm{halo}}}
\newcommand{\Om}{\ensuremath{\Omega_\mathrm{m}}}
\newcommand{\sig}{\ensuremath{\sigma_8}}
\newcommand{\LCDM}{\ensuremath{\Lambda\mathrm{CDM}}}
\newcommand{\Msun}{\ensuremath{\mathrm{M}_\odot}}

\defcitealias{bocquet24II}{Paper~II}

\begin{document}

\preprint{DES-2023-773}
\preprint{FERMILAB-PUB-23-521-PPD}

\title{SPT Clusters with DES and HST Weak Lensing. I. Cluster Lensing and Bayesian Population Modeling of Multi-Wavelength Cluster Datasets}

\author{S.~Bocquet}
\email{sebastian.bocquet@physik.lmu.de}
\affiliation{University Observatory, Faculty of Physics, Ludwig-Maximilians-Universit\"at, Scheinerstr. 1, 81679 Munich, Germany}
\author{S.~Grandis}
\affiliation{Universit\"at Innsbruck, Institut f\"ur Astro- und Teilchenphysik, Technikerstr. 25/8, 6020 Innsbruck, Austria}
\affiliation{University Observatory, Faculty of Physics, Ludwig-Maximilians-Universit\"at, Scheinerstr. 1, 81679 Munich, Germany}
\author{L.~E.~Bleem}
\affiliation{Argonne National Laboratory, 9700 South Cass Avenue, Lemont, IL 60439, USA}
\affiliation{Kavli Institute for Cosmological Physics, University of Chicago, Chicago, IL 60637, USA}
\author{M.~Klein}
\affiliation{University Observatory, Faculty of Physics, Ludwig-Maximilians-Universit\"at, Scheinerstr. 1, 81679 Munich, Germany}
\author{J.~J.~Mohr}
\affiliation{University Observatory, Faculty of Physics, Ludwig-Maximilians-Universit\"at, Scheinerstr. 1, 81679 Munich, Germany}
\affiliation{Max Planck Institute for Extraterrestrial Physics, Gie{\ss}enbachstr.~1, 85748 Garching, Germany}

\author{M.~Aguena}
\affiliation{Laborat\'orio Interinstitucional de e-Astronomia - LIneA, Rua Gal. Jos\'e Cristino 77, Rio de Janeiro, RJ - 20921-400, Brazil}

\author{A.~Alarcon}
\affiliation{Argonne National Laboratory, 9700 South Cass Avenue, Lemont, IL 60439, USA}

\author{S.~Allam}
\affiliation{Fermi National Accelerator Laboratory, P. O. Box 500, Batavia, IL 60510, USA}

\author{S.~W.~Allen}
\affiliation{Kavli Institute for Particle Astrophysics and Cosmology, Stanford University, 452 Lomita Mall, Stanford, CA 94305, USA}
\affiliation{Department of Physics, Stanford University, 382 Via Pueblo Mall, Stanford, CA 94305, USA}
\affiliation{SLAC National Accelerator Laboratory, 2575 Sand Hill Road, Menlo Park, CA 94025, USA}

\author{O.~Alves}
\affiliation{Department of Physics, University of Michigan, Ann Arbor, MI 48109, USA}

% ...

\collaboration{the DES and SPT Collaborations}
\noaffiliation

\date{Phys. Rev. D accepted 23 May 2024}

\begin{abstract}
We present a Bayesian population modeling method to analyze the abundance of galaxy clusters identified by the South Pole Telescope (SPT) with a simultaneous mass calibration using weak gravitational lensing data from the Dark Energy Survey (DES) and the {\it Hubble Space Telescope} (HST). We discuss and validate the modeling choices with a particular focus on a robust, weak-lensing-based mass calibration using DES data. For the DES Year 3 data, we report a systematic uncertainty in weak-lensing mass calibration that increases from 1\% at $z=0.25$ to 10\% at $z=0.95$, to which we add 2\% in quadrature to account for uncertainties in the impact of baryonic effects. We implement an analysis pipeline that joins the cluster abundance likelihood with a multi-observable likelihood for the Sunyaev-Zel'dovich effect, optical richness, and weak-lensing measurements for each individual cluster. We validate that our analysis pipeline can recover unbiased cosmological constraints by analyzing mocks that closely resemble the cluster sample extracted from the SPT-SZ, SPTpol~ECS, and SPTpol~500d surveys and the DES Year~3 and HST-39 weak-lensing datasets. This work represents a crucial prerequisite for the subsequent cosmological analysis of the real dataset.
\end{abstract}

\maketitle

\tableofcontents

\section{Introduction}

The abundance of massive dark-matter halos (and of the galaxy clusters they host) as a function of cosmic time -- the halo mass function -- depends sensitively on the cosmological parameters, and in particular the matter density \Om, the amplitude of fluctuations on 8~$h^{-1}$Mpc scales \sig, and the dark energy equation of state parameter $w$.
Therefore, measurements of the cluster abundance can be turned into a powerful cosmological probe e.g., \citep{haiman01}.
In practice, however, we cannot directly access the halo mass, but we can observe cluster properties that correlate with mass (so-called mass proxies, or simply observables).
These observables can be classified into three broad categories: i) optical and infrared properties of cluster member galaxies and of intra-cluster light, ii) properties of the gaseous intra-cluster medium (ICM), and iii) measurements of the effects of gravitational lensing.
So-called observable--mass relations then create the missing link between these measurements and the theoretical model for the halo mass function, and constraints on cosmology can be derived (see, e.g., \cite{allen11, pratt19} for reviews).

The South Pole Telescope (SPT) \citep{carlstrom11} detects galaxy clusters via the thermal Sunyaev-Zel'dovich effect (hereafter SZ) \cite{sunyaev&zeldovich72}, which is caused by cold photons from the cosmic microwave background (CMB) scattering with hot electrons in the ICM.
The SZ effect is a spectral distortion in the CMB radiation and is thus not affected by cosmic dimming.
Therefore, with its arcminute resolution that is well matched to the size of massive distant clusters, the SPT can detect clusters out to the highest redshifts at which they exist, probing a large range of cosmic times.
In practice, the SZ cluster candidates are confirmed by the presence of an overdensity of (cluster member) galaxies, which are also used to determine the cluster redshift.
The distinct signature of the SZ effect allows the construction of highly pure and complete cluster samples which are a strong foundation for cosmological analyses e.g., \citep{benson13, hasselfield13, bocquet15, dehaan16, planck18I, bocquet19}.
The strength of the SZ effect is given by the integrated ICM pressure and thus correlates tightly with the halo mass e.g., \citep{angulo12}.
However, due to our lack of sufficiently detailed knowledge about the properties of the ICM, the details (i.e., the parameters) of the SZ--mass relation cannot be predicted reliably.

The effects of gravitational lensing, on the other hand, are sourced by the entire matter distribution of a halo, and, on cluster-mass scales, are only mildly affected by the details of galaxy and ICM evolutions.
However, the typical measurement uncertainty is large and observations for many clusters need to be combined to obtain sufficient constraining power.
Then, lensing can offer a robust means of measuring halo masses, with well-understood control over systematic uncertainties (e.g., \cite{applegate14, bellagamba19, miyatake19, dietrich19, mcclintock19wl}, and review \cite{umetsu20}).

This is the first in a series of papers that aim at deriving cosmology and cluster astrophysics constraints by leveraging the overlap of the SPT survey and the Dark Energy Survey (DES) footprints.
Indeed, one of the science goals of the DES was to provide optical confirmation, redshifts, and weak-lensing measurements for SPT clusters.
Of the 5,200~deg$^2$ of SPT cluster surveys (combining SPT-SZ \citep{bleem15}, SPTpol~ECS \cite{bleem20}, and SPTpol~500d \citep{bleem23}), almost 3,600 are also covered by the DES.
In this overlap area, and up to cluster redshift of $z\sim1.1$, we now use DES data to confirm clusters in a statistically robust way by using information from random lines of sight to calibrate the frequency of (false) random associations.\footnote{We resort to the all-sky WISE survey at high cluster redshifts $z\gtrsim1.1$ that are beyond the reach of DES.}
This approach is implemented in the multi-component matched filter cluster confirmation  algorithm (MCMF) \cite{klein18} that has been applied to various cluster datasets \citep{hernandez-lang23, klein23_rass}, now including SPT \cite{bleem23, klein23_spt}.
The optical confirmation with MCMF allows us to use the SPT data to greater depth than outside of the DES overlap region, where we adopt an SZ-only selection scheme.
The resulting SPT cluster cosmology sample comprises 1,005 confirmed clusters above $z>0.25$.
We use DES weak-lensing data for 688 clusters up to $z<0.95$.
At high cluster redshifts $0.6<z<1.7$, the DES lensing dataset is supplemented with targeted measurements using the {\it Hubble Space Telescope} for 39 clusters (henceforth HST-39) \cite{schrabback18, schrabback21, zohren22}.

\begin{figure*}
  \includegraphics[width=\textwidth]{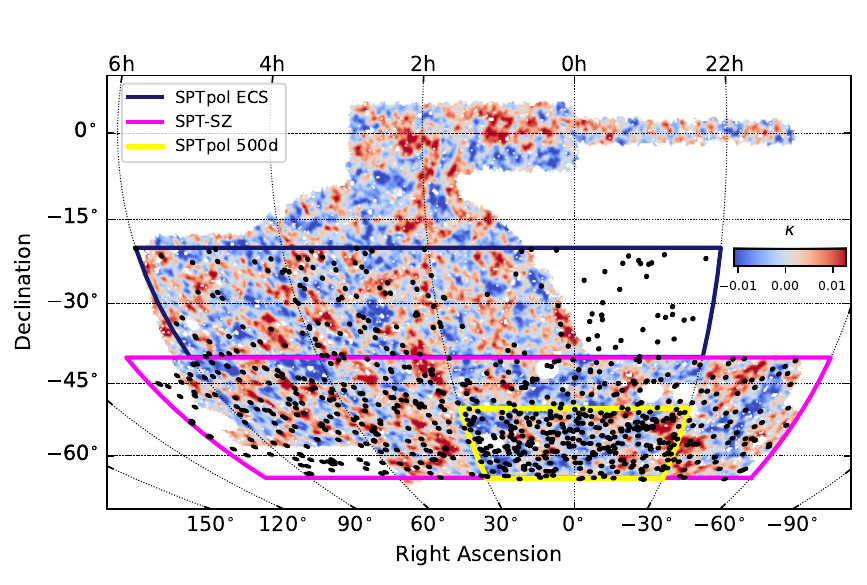}
  \caption{DES Y3 weak-lensing mass map, smoothed with a 0.5~deg Gaussian kernel \citep{jeffrey21}.
  We outline the main SPT cluster survey footprint (the SPTpol~ECS fields centered at R.A.~11 and 13~hrs, Dec.~$-25^\circ$ are not shown) and the subset of SPT clusters with $z<0.95$, for which we can measure the DES~Y3 lensing signal in the overlapping survey regions.
  The spatial density of clusters increases from SPTpol~ECS to SPT-SZ and SPTpol~500d due to the increasing survey depths, which enhance the sensitivity to lower-mass clusters that are more abundant.}
  \label{fig:map}
\end{figure*}

In this work, we describe the measurements and the analysis framework for the cluster cosmology analysis.
We focus on three key areas:
\begin{enumerate}
    \item We measure weak-lensing shear profiles of SPT clusters using the DES Year 3 (Y3) lensing dataset (see Fig.~\ref{fig:map}).
    To make these measurements useful in the calibration of the SPT observable--mass relation, we establish a framework that relates the lensing measurements to the underlying halo mass, following \citep{grandis21}.
    This framework requires additional measurement inputs such as the offset distribution of the observed cluster centers around which the shear profiles are measured, and a determination of the fraction of cluster member contamination, which tends to dilute the observed amount of shear.
    We perform these calibrations and combine them with synthetic mass maps from numerical simulations to obtain the complete weak-lensing mass calibration model.\footnote{Note that the measurements and model calibrations for the HST-39 lensing dataset were presented in separate publications and can be used without further modification \citep{schrabback18, schrabback21, zohren22}.}

    \item We present the analysis strategy and likelihood function for the cosmology analysis, which is designed to be simple yet robust, accounting for all relevant sources of biases and uncertainties.
    The framework is based on the notion that weak lensing can provide unbiased mass estimates within known and controlled uncertainties.
    Our approach is minimalist in the sense that we avoid making strong or unnecessary assumptions about, e.g., the parameters of the SZ--mass relation (in particular, we do not assume the clusters to be in hydrostatic equilibrium) or the properties of cluster member galaxies.
    Instead, using the weak-lensing data, we empirically calibrate the observable--mass relations.
    We use the multi-wavelength cluster data in a well-understood ``cluster-by-cluster'' likelihood framework that allows us to handle, e.g., possible correlations between the different cluster observables straightforwardly (to mitigate an effect often called ``selection bias'').

    \item We validate the analysis pipeline by analyzing mock catalogs.
    These synthetic datasets are drawn from our model and therefore include all known sources of systematic and statistical uncertainties.
    The analysis of several statistically independent mock realizations confirms that our likelihood framework and analysis pipeline are able to recover the known input values and are thus ready for the analysis of the real dataset.
\end{enumerate}

The paper is structured as follows.
We present the dataset in Sec.~\ref{sec:data}.
A summary of halo lensing theory can be found in Sec.~\ref{sec:cluster_lens_basics}.
We extract the DES weak-lensing measurements in Sec.~\ref{sec:DESmeasure}.
We discuss our model of the weak-lensing measurements in Sec.~\ref{sec:clusterlensmodel}, and the observable--mass relations in Sec.~\ref{sec:OMR}.
In Sec.~\ref{sec:likelihood}, we describe the likelihood function.
The validation of our analysis pipeline using mock data is presented in Sec.~\ref{sec:validation}.
We conclude with a summary in Sec.~\ref{sec:summary}.
The cosmological results of the joint SPT cluster and DES and HST weak-lensing dataset will be presented in a companion paper (see \cite{bocquet24II}; hereafter~\citetalias{bocquet24II}).

Throughout this paper, the (multivariate) normal distribution with mean $\vec\mu$ and (co)variance $\mathbf K$ is expressed as $\mathcal N(\vec\mu, \mathbf K)$.
When converting angles to distances, we adopt a fiducial flat $\Lambda$CDM cosmology with $\Om=0.3$ and $h=0.7$.
Halo masses $M_{200\mathrm{c}}$ refer to the mass enclosed within a sphere of radius $r_{200\mathrm{c}}$, within which the mean density is 200 times larger than the critical density $\rho_\mathrm{c}(z)$ at the cluster redshift $z$.
$M_{500\mathrm{c}}$ and $r_{500\mathrm{c}}$ are defined in an analogous way.

\section{Data}
\label{sec:data}

We construct a cluster catalog using SPT-SZ and SPTpol survey data and optical and infrared follow-up measurements.
We supplement the catalog with weak-lensing measurements from the DES and targeted observations using the HST.

\begin{table*}
\caption{\label{tab:SPTsample}
The SPT-SZ and SPTpol cluster surveys.
The SPTpol 500d footprint lies within SPT-SZ. Over the shared footprint, we use only the deeper 500d data.
The SPTpol ECS covers two separate regions of sky.
We quote the unmasked survey area used in the analysis, and the overlap with the unmasked part of the DES survey that can be used to confirm cluster candidates.
The number of clusters is for the cosmology sample used in this work.
The selection is summarized in Eqs.~(\ref{eq:select_noDES}) and (\ref{eq:select_DES}).}
\begin{ruledtabular}
\begin{tabular}{llccccc}
Survey & \multicolumn{2}{c}{Boundaries} & Depth (150~GHz) & Area & Area$\cap$DES & No. of clusters\\
&&& $[\mu$K-arcmin] & \multicolumn{2}{c}{[deg$^2$]} & with $z>0.25$\\
\colrule
SPTpol ECS & 10h $\leq$ R.A. $\leq$ 14h & $-30^\circ\leq$ Dec. $\leq-20^\circ$ & 30--40 & 541.8 & 0 & 41\\
SPTpol ECS & 22h $\leq$ R.A. $\leq$ 6h & $-40^\circ\leq$ Dec. $\leq-20^\circ$ & 25--39 & 1,986.3 & 1,421.6 & 166\\
SPT-SZ & 20h $\leq$ R.A. $\leq$ 7h & $-65^\circ\leq$ Dec. $\leq-40^\circ$ & 12--18 & 1,906.0 & 1,688.9 & 408\\
SPTpol 500d & 22h $\leq$ R.A. $\leq$ 2h & $-65^\circ\leq$ Dec. $\leq-50^\circ$ & 5.3 & 460.1 & 456.8 & 390\\
\colrule
Total &&&& 4,894.2 & 3,567.3 & 1,005
\end{tabular}
\end{ruledtabular}
\end{table*}

\subsection{The SPT Cluster Catalog}

We use a combination of the cluster catalogs from the SPT-SZ and SPTpol surveys, which cover a total of 5270~deg$^2$ of the southern sky.
Note that within the SPT-SZ survey footprint, the 500~deg$^2$ SPTpol~500d patch was re-observed to greater depth with SPTpol \citep{austermann12}.
The survey footprint is shown in Fig.~\ref{fig:map}.
Key features of the SPT cluster surveys are summarized in Table~\ref{tab:SPTsample}.

Over the entire survey region, the cosmology catalog only includes clusters above redshift $z>0.25$.
Objects at lower redshifts are excluded because, owing to the filtering applied to the SPT maps to remove atmospheric noise as well as increased noise contributions from the primary CMB, there is a strong evolution in the SPT selection function at low redshift.

Over the 1,327~deg$^2$ of the SPT survey that is not covered by DES, we apply the sample selection as a cut in the SPT detection significance $\xi>5$.
The resulting cluster candidate list has a purity $\gtrsim95\%$.
Cluster confirmation and redshift assignment are performed using targeted optical observations.
In particular, all SPT-SZ clusters and some SPTpol clusters were imaged in Sloan \textit{g}, \textit{r}, \textit{i}, and \textit{z} with the Parallel Imager for Southern Cosmology Observations (PISCO; \cite{stalder14}).
More details about the cluster samples can be found in the original catalog publications \cite{bleem15, bleem20}.

\begin{figure}
  \includegraphics[width=\columnwidth]{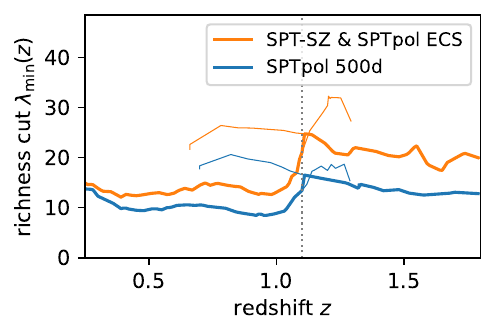}
  \caption{Sample selection threshold in optical richness $\lambda$ as a function of redshift. The threshold is empirically determined to ensure a redshift-independent sample purity.
  The dotted line marks redshift $z=1.1$, below which we use optical data from DES. We use WISE data above that redshift.
  Thin lines show the individual $\lambda_\mathrm{min}(z)$ for DES and WISE, and thick lines show the combination.
  The SPTpol~500d survey is significantly deeper than SPT-SZ and SPTpol~ECS and we thus apply a different sample selection threshold.
  }
  \label{fig:lambda_min}
\end{figure}

Over the 3,567~deg$^2$ of the SPT survey that is covered by DES (notably, the overlap region contains the SPTpol~500d survey), we confirm SPT cluster candidates and assign redshifts using MCMF \citep{klein18}.
In a first step, MCMF measures an optical richness (the sum of membership probabilities of all galaxies considered cluster members), the position of the optical center, and a redshift for each SPT detection.
In principle, an SPT detection with a corresponding richness and redshift measurement can be considered a confirmed cluster of galaxies.
However, given the abundance of galaxies on the sky, there is a chance that a small local overdensity of galaxies is erroneously associated with an SPT noise fluctuation.
Therefore, we also run MCMF on random locations in the DES footprint to determine the statistical properties of chance associations as a function of richness and redshift.
We then consider an SPT cluster candidate confirmed only if the probability of chance association is smaller than a given threshold.
We define this threshold such that the sample of confirmed detections has a purity $>98\%$.\footnote{This value is chosen such that the remaining level of contamination is within the shot noise of the total sample size.}
In practice, this threshold is implemented as a lower limit in richness $\lambda_\mathrm{min}$, and SPT detections are considered to be confirmed clusters if the measured richness exceeds this threshold.
We let the value of $\lambda_\mathrm{min}$ evolve with cluster redshift [$\lambda_\mathrm{min}(z)$, see Fig.~\ref{fig:lambda_min}] such that the resulting sample purity is constant at all redshifts.
More details can be found in the publications in which MCMF is applied to SPT-SZ and SPTpol~500d data \citep{klein23_spt, bleem23}.

The different SPT surveys have different depths and therefore, at fixed detection significance $\xi$, the cluster candidate lists have different purity levels.\footnote{Over a fixed survey area, and above a given $\xi$ cut, a given fixed number of false detections due to noise fluctuations is expected. However, a sufficiently deep survey would also detect many clusters (and thus obtain a highly pure candidate list), whereas an extremely shallow survey would detect no or very few clusters (and thus produce a low-purity candidate list that mostly consists of false detections).}
To keep the purity of the cluster samples extracted from the SPT sub-surveys roughly constant, we apply different cuts in the detection significance ($\xi>4.25$ for SPTpol~500d, $\xi>4.5$ for SPT-SZ, and $\xi>5$ for SPTpol~ECS), and we apply a different (lower) $\lambda_\mathrm{min}(z)$ to SPTpol~500d (blue in Fig.~\ref{fig:lambda_min}).

MCMF can use DES data only for clusters up to redshift $z\lesssim1.1$.
As can be seen in Fig.~\ref{fig:lambda_min}, $\lambda_\mathrm{min}$ starts to steeply increase above redshift $z\gtrsim1$, indicating that only high-richness clusters can be confirmed beyond this redshift.
To confirm clusters at higher redshifts, we also run MCMF on data from the Wide-field Infrared Survey Explorer (WISE) \cite{WISEobservatory} and repeat the analysis along random lines of sight to determine $\lambda_\mathrm{min}(z)$.
The thick lines in Fig.~\ref{fig:lambda_min} show the complete model for $\lambda_\mathrm{min}(z)$ with the transition from DES to WISE data at redshift $z=1.1$.

Over the full SPT footprint, there are 747 confirmed clusters above $\xi>5$ and $z>0.25$.
This sample is already twice as large as the one used in our previous cluster cosmology analyses based on SPT-SZ \citep{dehaan16, bocquet19}.
When we apply the (lower) cut in the SPT detection significance as discussed above, along with the confirmation using MCMF and $\lambda_\mathrm{min}(z)$, we obtain our fiducial cosmology sample of 1,005 confirmed clusters, a sample that is almost three times as large as the SPT-SZ cosmology catalog.

To summarize, outside of the DES overlap area (approximately 27\% of the total survey area), we select 110 clusters according to
\begin{equation}
  \label{eq:select_noDES}
  \begin{split}
    \xi&>5, \\
    z&>0.25.
  \end{split}    
\end{equation}
Within the DES overlap region (approximately 73\% of the survey area), the selection is
\begin{equation}
  \label{eq:select_DES}
  \begin{split}
    \xi&>4.25 \,/\, 4.5 \,/\, 5 \,\,(\text{500d / SZ / ECS}), \\
    \lambda&>\lambda_\mathrm{min}(z), \\
    z&>0.25,
  \end{split}
\end{equation}
with the $\xi$ limit and $\lambda_\mathrm{min}(z)$ chosen for the appropriate SPT survey.
This sub-sample contains 895 clusters.
In the cosmological analysis, we will explicitly model the full sample as selected according to Eqs.~(\ref{eq:select_noDES}) and (\ref{eq:select_DES}).

We note that in the redshift regime where we run MCMF both on DES and WISE, the two richness measurements are in reasonable agreement.
We attempted to further tune and correct the WISE richness measurements to exactly match those from DES, but were not successful.
Therefore, in the cosmological analysis, we will separately fit the DES richness--mass relation and the WISE richness--mass relation.
Finally, note that the moderate spatial resolution of WISE may limit its ability to confirm high-redshift clusters (see discussion in \citep{bleem23}).
To first order, such effects are absorbed by the WISE richness--mass relation and the intrinsic scatter in that relation.
Then, in the cosmological analysis in \citetalias{bocquet24II}, we will blindly compare the analysis of the $z<1$ cluster sample (which does not rely on WISE data) with the analysis of the full cluster sample.

\subsection{DES Y3 Weak-Lensing Data}

The 5,000~deg$^2$ DES was conducted in the $grizY$ bands using the Dark Energy Camera (DECam) \cite{flaugher15} on the 4\,m Blanco telescope at the Cerro Tololo Inter-American Observatory (CTIO) in Chile.
In this work we use data from the first three years of observations (DES~Y3), which cover almost the entire survey footprint.

\subsubsection{The Shape Catalog}

The DES~Y3 shape catalog \citep{gatti21} is constructed from the $r$, $i$, and $z$ bands using the \textsc{Metacalibration}
pipeline \citep{huff&mandelbaum17, sheldon&huff17}.
We refer the reader to other DES~Y3 publications for detailed information about the photometric dataset \citep{sevilla-noarbe21} and the point-spread function modeling \citep{jarvis21}.
After application of all source selection cuts, the DES~Y3 shear catalog contains about 100~million galaxies over an area of 4,143\;deg$^2$.
Depending on the exact definition, the effective source density is 5--6\;arcmin$^{-2}$.

\subsubsection{Source Redshifts and Shear Calibration}
\label{sec:DES_Pz}

\begin{figure}
  \includegraphics[width=\columnwidth]{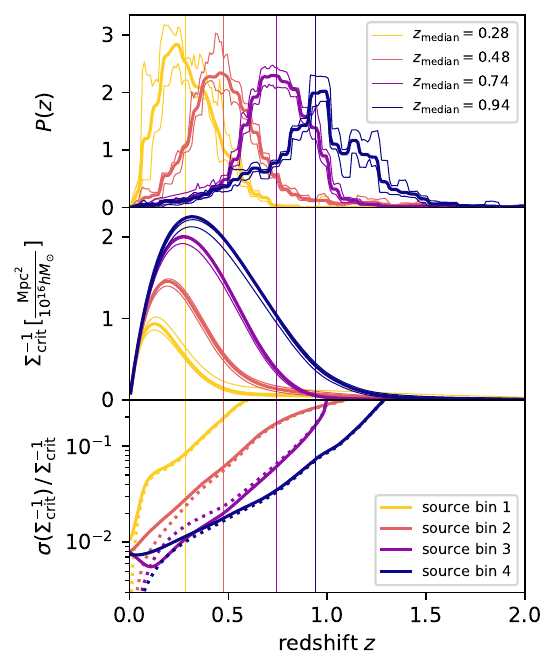}
  \caption{DES Y3 lensing source redshifts.
  Top panel: Mean and 5th/95th percentiles of the source redshift distributions in each tomographic bin.
  Middle panel: Lensing efficiency \invSigmac\ for each source bin, as a function of lens redshift.
  Bottom panel: (Systematic) Uncertainty in \invSigmac. Dotted lines show the contributions from photo-$z$ uncertainties, solid lines show the joint uncertainties from photo-$z$ and shear calibration. Throughout our analysis, we do not use source bin~1 due to its low lensing efficiency and the higher level of uncertainties in its calibration.
  }
  \label{fig:source_bins}
\end{figure}

We use the selection of lensing source galaxies in tomographic bins as defined and calibrated in \cite{gatti21, maccrann22, everett22, myles21} and employed in the DES 3x2~pt analysis \citep{DES_Y3_3x2pt}.
Source redshift distributions are estimated using self-organizing maps and the method is thus referred to as SOMPZ.
The final calibration accounts for the (potentially correlated) systematic uncertainties in source redshifts and shear measurements.
For each tomographic source bin, the mean redshift distribution (with amplitude scaled by factor $1+m$ to account for the multiplicative shear bias  $m$) is provided, and the systematic uncertainties are captured through 1,000 realizations of the distribution (see top panel of Fig.~\ref{fig:source_bins}).
Note that these uncertainties are correlated among the source bins; we account for this correlation in our analysis.

In addition to the tomographic bins and SOMPZ, we also use two individual galaxy photo-$z$ estimates, DNF \citep{devicente16} and BPZ \citep{benitez00}, when determinining the amount of cluster member contamination (see Sec.~\ref{sec:boost}).

\subsection{High-Redshift HST Weak-Lensing Data}
\label{sec:HSTdata}

To complement ground-based weak-lensing measurements, a sub-sample of 39 SPT clusters with redshifts $0.6<z<1.7$ were observed with the HST.
Additional photometric data were collected with VLT/FORS2 and Gemini-South/GMOS.
More details about the HST-39 dataset can be found in \cite{schrabback18, schrabback21, zohren22}.

\section{Galaxy Cluster Weak Lensing}
\label{sec:cluster_lens_basics}

Gravitational shear is induced by the matter density contrast in the lens plane. The tangential shear profile caused by a projected mass distribution $\Sigma(r)$ is
\begin{equation} \label{eq:gamma}
\gamma_\mathrm{t}(r) = \frac{\Delta\Sigma(r)}{\Sigma_\mathrm{crit}} = \frac{\langle\Sigma(<r)\rangle - \Sigma(r)}{\Sigma_\mathrm{crit}}.
\end{equation}
The critical surface mass density $\Sigma_\mathrm{crit}$ is defined as
\begin{equation} \label{eq:Sigmacrit}
    \invSigmac(\mathrm{source}, \mathrm{lens}) = \frac{4\pi G}{c^2} \frac{D_\mathrm{l}}{D_\mathrm{s}} \times 
    \mathrm{max}\left[0, D_\mathrm{ls} \right],
\end{equation}
where $c$ is the speed of light, $G$ the gravitational constant, and the $D_i$ are angular diameter distances, where $l$ denotes the lens and $s$ denotes the source.
When the source is not behind the lens, \invSigmac\ and the shear $\gamma$ vanish.

The observable quantity is the reduced tangential shear
\begin{equation} \label{eq:gt}
    g_\mathrm t(r) = \frac{\gamma_\mathrm{t}}{1-\kappa}(r)
\end{equation}
with the convergence $\kappa(r) = \Sigma(r)/\Sigma_\mathrm{crit}$.

The inverse critical surface mass density \invSigmac\ plays a central role in the lensing analysis because it acts as a lensing efficiency that modulates the strength of the observed shear signal given a particular lens mass. To compute \invSigmac, the lens redshift and the redshift distribution of source galaxies need to be known, and uncertainties in the calibration of that distribution propagate into uncertainties on \invSigmac\ (see middle and bottom panels of Fig.~\ref{fig:source_bins}). 
Residual uncertainties in the shear calibration further affect the relation in Eq.~(\ref{eq:gt}).
In our analysis of DES~Y3 lensing data, we account for both of these effects as discussed in Sec.~\ref{sec:DES_Pz}.

Cluster lensing is measured along cluster sightlines, where cluster member galaxies -- which are not sheared by their host halo -- can potentially contaminate the sample of source galaxies.
This cluster member contamination biases the measured shear low, and it is particularly important when the galaxy redshifts are estimated from broad-band photometry.
We characterize this contamination in Sec.~\ref{sec:boost} and account for it in Sec.~\ref{sec:DESmodel}.

Finally, the massive dark matter halos that host galaxy clusters are complex objects that are embedded in the large-scale structure. The variety of halo profiles, their correlation with neighboring structures, and uncertainties in the observationally determined halo centers all need to be modeled and accounted for.
We will address these points in Secs.~\ref{sec:miscentering} and \ref{sec:clusterlensmodel}.

\section{DES Weak-Lensing Measurements}
\label{sec:DESmeasure}

We extract DES Y3 weak-lensing data products for SPT clusters and quantify the relevant systematic and statistical uncertainties. In short:
\begin{enumerate}
    \item We define rescaling factors for the shear in each tomographic source bin. These factors depend on cluster redshift and allow us to optimally combine the source bins for each lens;
    \item We extract the tangential shear profile and source redshift distribution for each cluster in our sample;
    \item We determine the miscentering distributions of the observationally determined cluster centers;
    \item We estimate the cluster member contamination of the measured shear signal.
\end{enumerate}
These steps are described in detail in the following subsections.
In Sec.~\ref{sec:DESmodel} we fold these measurements into a condensed model that is then implemented in the cosmology analysis pipeline.

\subsection{Tomographic Source Bins}
\label{sec:tomo_bins}

We select lensing source galaxies based on four tomographic bins (see Fig.~\ref{fig:source_bins}), following the DES 3x2~pt analysis \citep{DES_Y3_3x2pt}.
Assuming a fiducial flat \LCDM\ cosmology with $\Om=0.3$, we compute the average lensing efficiency for each bin $b$
\begin{equation} \label{eq:meaninvSigmac}
    \langle\invSigmac_b\rangle(z_\mathrm{lens}) = \int \dif z_\mathrm{s} \,P_b(z_\mathrm{s}) \,\invSigmac(z_\mathrm{s}, z_\mathrm{lens})
\end{equation}
using the mean source redshift distribution $P_b(z_\mathrm{s})$, and
as shown in the middle panel of Fig.~\ref{fig:source_bins}.

For a given lens, we only use those source bins for which the median source redshift is larger than the lens redshift, to avoid the regime where the analysis would be highly sensitive to the accurate calibration of the high-redshift tails of the redshift distributions.
The median redshifts of the four bins are $z_\mathrm{median}=0.285, 0.476, 0.743, 0.942$, as indicated by vertical lines in Fig.~\ref{fig:source_bins}.
In Appendix~\ref{sec:lens_eff_ratio}, we discuss that applying more aggressive limits (i.e., use the source bins up to higher lens redshifts) leads to unexpected trends in the redshift evolution of the lensing efficiencies, and we thus prefer to apply the fiducial selection described above.
In principle, for clusters with redshifts $0.25<z<0.28$ we could use lensing sources from bin~1. However, because of the relatively high level of uncertainty in its calibration, we discard bin~1 altogether.
In summary, we do not use source bin~1, and we use bin~2 for lenses with $z<0.47$, bin~3 for lenses with $z<0.74$, and bin~4 for lenses with $z<0.95$.
With these cuts, we extract over 99\% of the total statistical constraining power (signal-to-noise ratio) of the lensing dataset.

\subsection{Shear Measurements and Source Redshift Distributions}
\label{sec:shear_meas}

In DES~Y3, the lensing shear is extracted using \textsc{Metacalibration} \citep{huff&mandelbaum17, sheldon&huff17}.
The Taylor expansion of the observed source ellipticity $\vec e$ given an applied amount of shear $\vec \gamma$ yields,
\begin{equation}
  \begin{split}
    \vec e =& \vec e|_{\gamma=0} + \frac{\partial \vec e}{\partial \gamma}\big|_{\gamma=0} ~\vec \gamma + ... \\
 \equiv& \vec e|_{\gamma=0} + \mathbf R_{\vec\gamma} ~\vec \gamma+...
  \end{split}
\end{equation}
with the shear response $\mathbf R_{\vec\gamma}$.
Since the mean unsheared ellipticity vanishes ($\langle\vec e|_{\gamma=0}\rangle=0$) we obtain an estimator for the average shear,
\begin{equation}
    \langle \vec \gamma\rangle = \langle \mathbf R_{\vec\gamma} \rangle^{-1} \langle \vec e \rangle.
\end{equation}
The shear response is computed from artificially sheared shape catalogs. In practice, we use a smooth shear response estimator $R_\gamma$ (see Sec.~4.3 in Ref.~\citep{gatti21}).
Additionally, a selection response $R_\mathrm{sel}$ accounts for the fact that lensing sources are selected based on their (intrinsically) sheared observations.
We determine a single value of $R_\mathrm{sel}$ for the entire sample of cluster lensing sources:
\begin{equation}
    R_\mathrm{sel} \approx \frac12\frac{\langle e_1\rangle^{S+} - \langle e_1\rangle^{S-} + \langle e_2\rangle^{S+} - \langle e_2\rangle^{S-}}{\Delta \gamma},
\end{equation}
where $e_1, ~e_2$ are the ellipticities along the Cartesian coordinate axes, and where the superscripts $S+$ and $S-$ indicate that artificial shear of $+0.01$ and $-0.01$ is applied (and thus $\Delta\gamma=0.02$).

Notionally, a simple estimator for the tangential shear can be defined by averaging over all sources $i$ in all source bins $b$ as follows:
\begin{equation}
    g_\mathrm{t,\,preliminary} = \frac{\sum_{b=2,3,4} \sum_i e_{\mathrm t,b,i} \,w^\mathrm{s}_i }{\sum_{b=2,3,4}\sum_i w^\mathrm{s}_i\, (R_\gamma + R_\mathrm{sel})},
\end{equation}
with source weights $w^\mathrm{s}_i$ (corresponding to the inverse variance in the measured ellipticity, accounting both for the intrinsic variance of shapes and for measurement uncertainties) and shear and selection response $R_\gamma$ and $R_\mathrm{sel}$.

In practice, we refine the estimator by accounting for the fact that the lensing efficiency changes between the source bins, and with it, the amplitude of the observed shear.
The preliminary estimator averages over data that do not have a common mean, thereby artificially increasing the variance in the recovered estimate.
To avoid this effect, we rescale the ellipticities in each bin $b$ by a factor of
\begin{equation}
  \label{eq:f_rescale}
  f_{\mathrm{rescale,bin}\,b}(z_\mathrm{lens}) = \langle\Sigma_{\mathrm{crit,bin}4}^{-1}\rangle(z_\mathrm{lens})/\langle\Sigma_{\mathrm{crit,bin}b}^{-1}\rangle(z_\mathrm{lens})
\end{equation}
and divide the weights in bin $b$ by $f_{\mathrm{rescale,bin}\,b}^2(z_\mathrm{lens})$.
By definition, the shear in bin~4 remains unchanged, and the shear in the other source bins is enhanced.\footnote{The noise in the other bins is enhanced accordingly, so that the signal-to-noise ratio per bin remains constant. In other words, our estimator does not alter the signal-to-noise ratio per source bin, but it does increase the signal-to-noise ratio of the combined measurement.}
The estimator we employ in our analysis then is
\begin{equation}
    g_\mathrm{t} = \frac{\langle\Sigma_{\mathrm{crit,bin}\,4}^{-1}\rangle(z_\mathrm{lens}) \sum_{b=2,3,4} \langle\Sigma_{\mathrm{crit,bin}\,b}^{-1}\rangle(z_\mathrm{lens}) \sum_i e_{\mathrm t,b,i} \,w^\mathrm{s}_i }{\sum_{b=2,3,4} \left(\langle\Sigma_{\mathrm{crit,bin}\,b}^{-1}\rangle(z_\mathrm{lens})\right)^2 \sum_i w^\mathrm{s}_i \,(R_\gamma + R_\mathrm{sel})}.
\end{equation}
Note that, as defined in Eqs.~(\ref{eq:Sigmacrit}) and (\ref{eq:meaninvSigmac}), the $\langle\Sigma_\mathrm{crit}^{-1}\rangle$ used in the estimator are computed in our reference cosmology.
In the cosmological analysis, model shear profiles are computed using $\Sigma_\mathrm{crit}^{-1}$ evaluated for each tomographic bin for the cosmological parameters of the evaluation, and then the model shear profiles are combined into a single profile using the (fixed, non-cosmology dependent) rescaling factors.
We estimate the uncertainties on the shear profiles by bootstrap resampling the individual sources.
Because the bootstrap uncertainties are noisy, especially for the inner radial bins with few sources, we determine the characteristic shape noise to be 0.37 and assign Gaussian uncertainties of width $0.37/\sqrt{N_\mathrm{source}}$.\footnote{The shape noise we determine is larger than quoted in the presentation of the shape catalog \citep{gatti21} because of the rescaling we apply to the ellipticities [see Eq.~(\ref{eq:f_rescale})]. The resulting signal-to-noise ratio remains unchanged.}

We also experimented with a more sophisticated estimator where $f_\mathrm{rescale}$ is additionally multiplied with the source fraction (unity minus the cluster member contamination, see Sec.~\ref{sec:boost}) in the respective source bin.
This scheme was meant to further increase the signal-to-noise ratio in the measured shear profiles.
However, in practice, the improvements were modest and therefore, we do not apply this extra analysis step.

For each cluster, we also measure the lensing source redshift distribution as
\begin{equation} \label{eq:source_redshift_distribution}
    P(z_\mathrm{s}) = \frac{\sum_{b=2,3,4} \left(\langle\Sigma_{\mathrm{crit,bin}\,b}^{-1}\rangle(z_\mathrm{lens})\right)^2 P_{b}(z_\mathrm{s}) \sum_i R_\gamma w^\mathrm{s}_i}{\sum_{b=2,3,4} \left(\langle\Sigma_{\mathrm{crit,bin}\,b}^{-1}\rangle(z_\mathrm{lens})\right)^2 \sum_i R_\gamma w^\mathrm{s}_i}
\end{equation}
where $P_{b}(z_\mathrm{s})$ is the mean source redshift distribution in each tomographic bin.
The distributions $P(z_\mathrm{s})$ will be used in the cosmological analysis to compute $\langle\invSigmac\rangle$ [in a way that is analogous to Eq.~(\ref{eq:meaninvSigmac})].

\begin{figure}
  \includegraphics[width=\columnwidth]{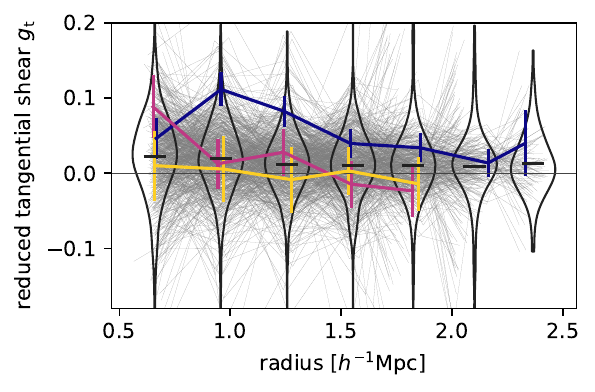}
  \caption{DES Y3 shear profiles of SPT clusters.
  The individual profiles (thin gray lines) are noisy, but on average, a positive shear signal is measured (dark violins show the distribution of measurements, dark horizontal marks show the mean).
  Colored lines show three example clusters with high (purple), average (pink), and low signal-to-noise ratios (yellow) in the lensing profile measurements.
  }
  \label{fig:shear_profiles}
\end{figure}

\begin{figure}
  \includegraphics[width=\columnwidth]{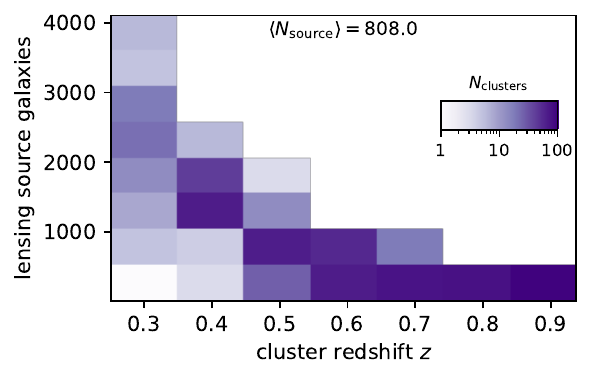}
  \caption{Number of DES Y3 lensing source galaxies per SPT cluster, as a function of cluster redshift.
  The number drops toward high cluster redshifts for several reasons; we use fewer source bins for high-redshift clusters, high-redshift clusters appear smaller on the sky, and additionally, we apply an outer radial cut of $3.2/(1+z)\,h^{-1}$Mpc.
  }
  \label{fig:N_source}
\end{figure}

We extract two sets of shear profiles and source redshift distributions for each cluster, using either the SPT position or the position determined from optical data using the MCMF algorithm.
We use sources between projected distances of 500~$h^{-1}$kpc and $3.2/(1+z_\mathrm{cluster})\,h^{-1}\mathrm{Mpc}$, where these regions are calculated within our fiducial cosmology.
We measure $R_\mathrm{sel} = -0.0023$ for optical centers and $R_\mathrm{sel} = -0.0025$ for SPT centers.
Compared to the typical shear response $R_\gamma\sim0.66$, the selection response thus plays a minor role.
We use annuli that are linearly spaced with $\Delta r = 0.3\,h^{-1}\mathrm{Mpc}$.
For each cluster and for each cluster radial bin, we measure the reduced shear and extract a source redshift distribution.
The tangential shear measurements are illustrated in Fig.~\ref{fig:shear_profiles}.
In total, for the optical centers, we extract lensing data around 688~clusters from a total of 555,912~sources, with an average of 808~sources per cluster (see Fig.~\ref{fig:N_source}).
The individual shear profile measurements are quite noisy as shown in the middle panel of the figure. 
In the analysis, we will combine lensing information for hundreds of clusters to obtain precise mass calibration constraints.

\subsection{The Miscentering Distributions for the Weak-Lensing Measurements}
\label{sec:miscentering}

Two sets of centers are available for all DES weak-lensing measurements; mm-wave SZ centers as measured by the SPT and optical centers extracted from the optical imaging using the MCMF algorithm.\footnote{MCMF adopts the brightest cluster galaxy (BCG) as the center if it is within 250~kpc of the cluster position determined by SPT, else, the position of the peak of the galaxy density map is used.}
Since no observationally determined position is a perfect tracer of the true halo center, the effect of miscentering must be accounted for in the lensing analysis.  Note that in the simulation-based models for the halo mass function and for cluster lensing, the potential minima are adopted as the true halo centers.

We set up the DES lensing analysis such that we can use either the SPT centers or the optical centers.
Using the real data, we will blindly compare the cosmological constraints obtained from the two sets of centers in \citetalias{bocquet24II}.
In this section, we calibrate the offset distributions between the observed positions and the underlying halo center, assumed to be the projected position of the minimum in the halo potential.

Some clusters in the lensing sample are heavily affected by masking in the DES data.
Because masking out the center strongly affects the determination of the optical center, we discard problematic clusters from the miscentering analysis and from the lensing sample.
In the end, we discard 10~clusters for which more than 1/3 of the area contained within a 1~arcmin radius of the center is masked.
In the case of SPT centers, no clusters are excluded.

\subsubsection{Fitting the SPT--optical offset distribution}
\label{sec:SPT-optical_miscenter}

We model the intrinsic miscentering distributions (SZ--true and optical--true) as mixtures of well-centered and miscentered cluster positions, each described by a Rayleigh distribution $\mathcal R$ of scale $\sigma$:
\begin{equation}
  \label{eq:miscenter}
  \begin{split}
    \sigma_i &= \sigma_{i,0} \left(\frac\lambda{60}\right)^{1/3} \text{for } i \in \{0,1\},\\
    P_\mathrm{offset}(r) &= \rho \, \mathcal R(r, \sigma_0) + (1-\rho) \, \mathcal R(r, \sigma_1).
  \end{split}
\end{equation}
The mixture weights $\rho$ and $1-\rho$ must be between 0 and 1 which we enforce by applying a uniform prior $\rho~\sim\mathcal U[0,1]$.
The scale $\sigma$ is commonly described as a function of $r_{500\mathrm{c}}$ e.g., \cite{lin04, song12, saro15, gupta17_sim, bleem20}. Given the approximately linear scaling of richness with mass e.g., \cite{mcclintock19wl, capasso19, bleem20}, we adopt a scaling of $\sigma$ with $\lambda^{1/3}$.

In addition to the intrinsic SZ miscentering distribution, the observed SPT centers are affected by noise and the telescope's positional uncertainty.
We model these effects as another Rayleigh distribution of scale
\begin{equation}
    \sigma_\text{SPT}^2 = \frac{\theta_\mathrm{beam}^2 + (\kappa_\mathrm{SPT}\theta_\mathrm{c})^2}{\xi^2} + \sigma_\mathrm{astrom.}^2 \label{eq:SPT_pos_unc}
\end{equation}
with the cluster detection significance $\xi$, the filter scale $\theta_\mathrm{c}$, an astrometric uncertainty $\sigma_\mathrm{astrom.}=5''$, and the fit parameter $\kappa_\mathrm{SPT}$ that is of order unity.
The effective SPT beam is $\theta_\mathrm{beam}=1.3'$.
We neglect the measurement uncertainty on the centers determined from the optical DES data.

The observed SZ--optical distribution is the convolution of the offset distribution between the true halo center and the SZ center, the offset distribution between the true halo center and the optical center, and the SPT positional uncertainty.
With this approach, we make the underlying assumption that the SZ and optical offsets for a given cluster are independent,
\begin{equation}
    P_\mathrm{SZ-optical}(r) = \bigl(P_\mathrm{SPT} * P_\mathrm{SZ} * P_\mathrm{optical}\bigr)(r).
\end{equation}
The log-likelihood function for the full cluster sample is
\begin{equation}
    \ln \mathcal L = \sum_i \ln P_\mathrm{SZ-optical}(r_i) + \mathrm{const.}
\end{equation}
with a measured offset $r_i$ for each cluster.
We validate our miscentering analysis code by analyzing mock datasets. We create these by taking the observed distribution of clusters in $\lambda-z$ space and then drawing mock offsets according to our model.

The recovered parameters of the miscentering model are summarized in Table~\ref{tab:miscentering}.
Note that the posterior distribution is mildly bimodal (see blue contours in Fig.~\ref{fig:miscenter_GTC}), because our model with two unknown offset distributions is very flexible.
When incorporating the miscentering model into our lensing modeling framework, we will use the full posterior distribution to correctly handle the bimodality.
In Appendix~\ref{sec:X-ray_miscenter}, we show that our constraints could be further refined by also using X-ray center positions.
However, that analysis requires extra assumptions, and we base our cosmology analysis on the observed SPT--optical offsets.

\begin{table}
\caption{\label{tab:miscentering}
Parameters of the SZ and optical miscentering distributions (mean and 68\% credible interval, one-sided limits are for the 95\% credible interval).}
\begin{ruledtabular}
\begin{tabular}{ll}
Parameter & Constraint\\
\colrule
$\rho_\mathrm{SZ}$ & $0.88^{+0.12}_{-0.06}$\\
$\sigma_{\mathrm{SZ},0}\, [h^{-1}\mathrm{Mpc}]$ & $0.007^{+0.002}_{-0.007}$\\
$\sigma_{\mathrm{SZ},1}\, [h^{-1}\mathrm{Mpc}]$ & $0.174^{+0.050}_{-0.113}$\\
$\kappa_\mathrm{SPT}$ & $0.92^{+0.14}_{-0.12}$\\
\colrule
$\rho_\mathrm{opt}$ & $0.89^{+0.11}_{-0.06}$\\
$\sigma_{\mathrm{opt},0}\, [h^{-1}\mathrm{Mpc}]$ & $0.007^{+0.002}_{-0.007}$\\
$\sigma_{\mathrm{opt},1}\, [h^{-1}\mathrm{Mpc}]$ & $0.182^{+0.038}_{-0.112}$
\end{tabular}
\end{ruledtabular}
\end{table}

As a cross-check, we allow for additional flexibility by allowing all parameters in Eq.~(\ref{eq:miscenter}) to evolve with redshift and richness.
However, in this more flexible analysis, all evolution parameters are consistent with no evolution, and we thus keep our simplified model.

\begin{figure}
  \includegraphics[width=\columnwidth]{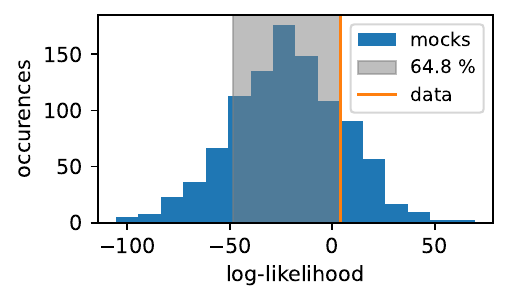}
  \caption{Goodness of fit of our miscentering model.
  We create 1,000~mocks of the SZ--optical offset distribution using the mean recovered fit parameters. We then compute the log-likelihood for each mock and for the real data, again using the mean recovered fit parameters.
  The log-likelihood of the real data is contained within the 64.8~\% interval (probability to exceed 0.352, corresponding to $0.9\sigma$) and we conclude that the model is adequately describing the data.}
  \label{fig:miscenter_validation}
\end{figure}

We further validate our model and the recovered constraints by drawing 1,000~mock realizations of the SZ--optical miscentering distribution using the mean recovered parameter values and the observed distribution of richness, $\theta_c$, and $\xi$.
We then compute the log-likelihood of the data and of each mock dataset given the mean recovered parameter values.
Within the distribution of log-likelihoods of the mocks, the log-likelihood of the real data has a probability to exceed of 0.352 ($0.9\sigma$) and we conclude that our model is able to adequately describe the data (see Fig.~\ref{fig:miscenter_validation}).

\begin{figure}
  \includegraphics[width=\columnwidth]{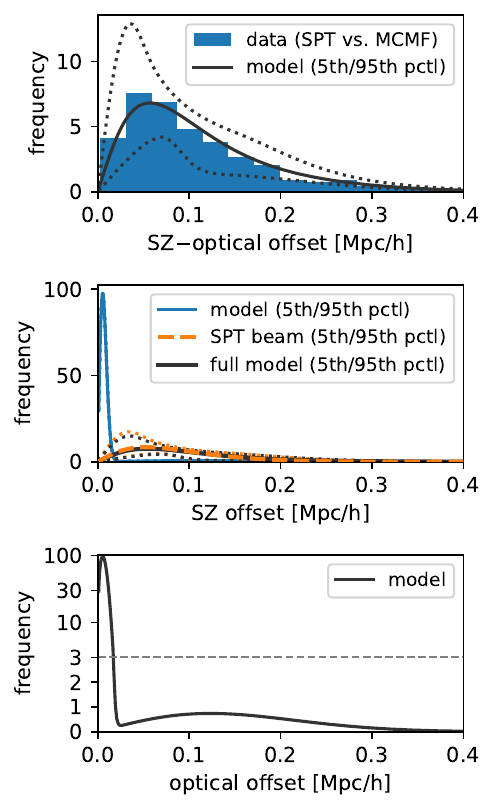}
  \caption{Offset distributions (SZ--optical, SZ--true, and optical--true).
  The top panel also shows the measured SPT--MCMF offset distribution which is well described by the model.
  The middle panel shows that the SZ offsets as measured by the SPT are dominated by the effect of the beam.}
  \label{fig:miscenter}
\end{figure}

In the top panel of Fig.~\ref{fig:miscenter}, we show the model predictions for the SZ--optical offset distribution along with the observed offsets.
In the middle panel, we show the SZ offset distribution and the contribution from the SPT beam (i.e., measurement uncertainty), which is the dominant source of SZ miscentering.
Finally, the bottom panel shows the optical offset distribution. Our model suggests that 81\% of the cluster population is well centered, whereas the remaining clusters show a typical offset of about 0.2~$h^{-1}$Mpc.
This result is broadly consistent with other analyses that used data from DES and precursor optical datasets e.g., \cite{lin04, zhang19, kelly23}.

\subsubsection{Comparison with Numerical Simulations}
\label{sec:miscenter_sims}

\begin{figure}
  \includegraphics[width=\columnwidth]{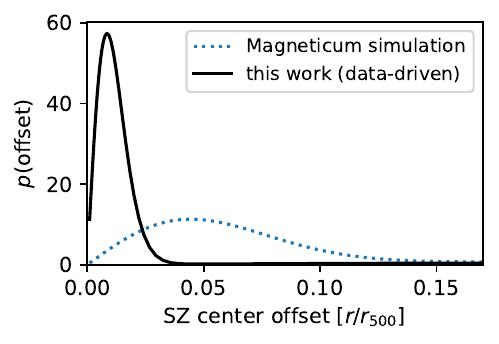}
  \caption{Offset distribution of the SZ cluster center with respect to the true halo center. We compare our data-driven result with the distribution obtained from light cones generated from the \textit{Magneticum} hydrodynamical simulation \citep{gupta17_sim}. Our empirically determined model (using the observed offsets between SPT centers and optical centers) is not well-described by the simulation and we thus do not use the simulation-based miscentering distribution in our analysis.}
  \label{fig:miscenter_sim}
\end{figure}

In the above, we establish the SZ and optical miscentering distributions empirically.
We now compare these results with measurements extracted from numerical simulations.
In Fig.~\ref{fig:miscenter_sim}, we show our model-inferred SZ centering distribution in units of $r/r_{500\mathrm{c}}$.
Our miscentering model discussed above is in physical units, and we convert to $r/r_{500\mathrm{c}}$ by using a fiducial value of $r_{500\mathrm{c}}$ for each cluster. The plot then shows the sample average.

The \textit{Magneticum} simulation\footnote{\url{http://www.magneticum.org/index.html}} has been previously used to infer the SZ miscentering distribution \cite{gupta17_sim}.
As a cross-check, we use the same mm-wave light cone map and add realizations of the CMB background, atmospheric foregrounds, and run the SPT cluster detection pipeline. We derive essentially the same miscentering distribution as presented in \cite{gupta17_sim}, confirming their result.
Figure~\ref{fig:miscenter_sim} suggests that the simulated light cone constructed from \textit{Magneticum} overestimates the amount of SZ miscentering. This is not a new realization; The SZ miscentering distribution by \cite{gupta17_sim} is in good agreement with measurements of the SZ--optical miscentering \citep{saro15, bleem20}. This would imply either that optical miscentering is negligible (which is ruled out by observations), or that \textit{Magneticum} overestimates the amount of SZ miscentering.
Other likely explanations are related to artifacts due to the construction of the light cone, or that there is significant correlation between SZ and optical miscentering.

Comparisons with more numerical simulations are needed to reach definitive conclusions on the observed mismatch.
We emphasize that these simulations need to be processed and analyzed as the SPT data would to make meaningful comparisons.
In this analysis, we proceed with our data-driven miscentering model and leave a more exhaustive comparison with simulations to future work.
In \citetalias{bocquet24II}, we will investigate the cosmological impact of using either the optical centers or the SPT centers, or larger radial scales to verify that our miscentering model is sufficiently robust.

\subsection{Cluster Member Contamination of the Lensing Source Galaxy Sample}
\label{sec:boost}

The sample of lensing source galaxies along a cluster line of sight is in general contaminated by cluster member galaxies because we measure redshifts using broad-band photometry with relatively large statistical uncertainties.
These galaxies, which are not sheared by their host halo, bias the measured weak-lensing signal low.
To quantify the fractional contamination by cluster member galaxies, we closely follow the methodology described in \cite{paulus_thesis}.
In that work, the method of $P(z)$ decomposition e.g.,~\cite{gruen14, dietrich19, stern19, varga19} is applied in a cluster-by-cluster analysis context like ours, i.e., without stacking.
In this approach, the redshift distribution of source galaxies is modeled as the weighted sum of an uncontaminated field component $P_\mathrm{field}(z)$, and a component of the contaminants $P_\text{cl}(z)$.
In the original analysis, the focus was on DES Y1 data; we update aspects of the analysis and apply it to the DES~Y3 dataset.

The radial dependence of the contaminants is modeled with a Navarro-Frenk-White profile (NFW) \citep{navarro97}, which we normalize to unity at $r=1\,h^{-1}\mathrm{Mpc}$. To approximately account for the effect of miscentering, the profile is modified to remain constant within the miscentering radius $R_\mathrm{mis}$, which we define as
\begin{equation}
  \label{eq:Rmis}
  R_\mathrm{mis} = \sqrt\frac\pi2 \bigl(\rho \sigma_0 + (1-\rho)\sigma_1\bigl)
\end{equation}
with the mean miscentering parameters $\rho$ and $\sigma$ from Sec.~\ref{sec:miscentering}.
We model the scale radius $r_s$ of the NFW profile as a function of cluster richness as
\begin{equation}
    r_s = \frac{(\lambda/60)^{1/3}}{10^{c_\lambda}}
\end{equation}
with the free parameter $c_\lambda$.
We allow for a power-law dependence of the cluster member contamination with cluster richness.
The dependence with cluster redshift $z_\mathrm{cl}$ is complicated \citep{paulus_thesis}, and we allow for considerable freedom. The full model reads
\begin{equation}
  \label{eq:boost_model}
  \begin{split}
    A(R, z_\mathrm{cl}, \lambda) =& \Sigma_\mathrm{NFW}(R, r_s)/\Sigma_\mathrm{NFW}(1\,h^{-1}\mathrm{Mpc}, r_s) \\
&\times \exp\left(A_\infty + \sum_i A_i \exp\left(-\frac12\frac{(z_\mathrm{cl}-z_i)^2}{\rho_\mathrm{corr}^2}\right)\right)\\
&\times (\lambda/60)^{B_{f_\mathrm{cl}}}
  \end{split}
\end{equation}
with the array of redshifts
\begin{equation}
    z_i \in \{0.2 , 0.28, 0.36, 0.44, 0.52, 0.6 , 0.68, 0.76, 0.84, 0.92, 1\}.
\end{equation}
The fractional cluster member contamination then is
\begin{equation}
\label{eq:f_cl}
f_\mathrm{cl}(R, z_\mathrm{cl}, \lambda) = A(R, z_\mathrm{cl}, \lambda)/\bigl(1+A(R, z_\mathrm{cl}, \lambda)\bigr).
\end{equation}
This functional form ensures that $0\leq f_\mathrm{cl}\leq1$ for any positive value of $A(R,z_\mathrm{cl},\lambda)$. 

We model the redshift distribution of source galaxies (with source redshift $z_\mathrm{s}$) as the weighted sum of the field distribution and a cluster member component, which is modeled as a Gaussian distribution of width $\sigma_z$ and that is offset from the cluster redshift by $z_\mathrm{off}$:
\begin{equation}
  \begin{split}
    P(z_\mathrm{s}, R, z_\mathrm{cl}, \lambda) =& f_\mathrm{cl}(R, z_\mathrm{cl}, \lambda) \, \mathcal N\Bigl(z_\mathrm{s} - (z_\mathrm{cl}+z_\mathrm{off}), \sigma_z^2\Bigr) \\
    &+ \bigl(1-f_\mathrm{cl}(R, z_\mathrm{cl}, \lambda)\bigr)\, P_\mathrm{field}(z_\mathrm{s}), \\
    z_\mathrm{off}(z) =& z_\mathrm{off,0} + (z_\mathrm{cl}-0.5) \, z_\mathrm{off,1}, \\
    \sigma_z(z) =& \sigma_{z,0} + (z_\mathrm{cl}-0.5) \, \sigma_{z,1}.
  \end{split}
\end{equation}
With no prior knowledge of the possible evolution of the offset $z_\mathrm{off}$ and width $\sigma_z$, we allow both to evolve linearly with redshift.\footnote{When sampling the likelihood, we reject parameter combinations of $\sigma_{z,0}$ and $\sigma_{z,1}$ that would result in $\sigma_z<0$.}

For each cluster, the likelihood for the observed sources in each bin in radius and source redshift is
\begin{equation}
\label{eq:boost_likelihood}
\ln \mathcal L_\mathrm{cluster} = \sum_i w_i \ln\bigl(P(z_{\mathrm{s},i}, R_i, z_\mathrm{cl}, \lambda)\bigr) + \mathrm{const.}
\end{equation}
where $w_i$ are the lensing weights of the source galaxies.
To correctly normalize the likelihood, we normalize the weights $w_i$ such that the mean weight equals the mean shear response $R_\gamma$.
In other words, the typical source galaxy contributes to the total likelihood with a weight of $\langle R_\gamma\rangle\approx0.66$.

Our model has considerable freedom along the redshift axis, see Eq.~(\ref{eq:boost_model}).
Based on the model parameter $\rho_\mathrm{corr}$, we impose a certain degree of smoothness.
For each pair of amplitudes $A_i$ and $A_j$ and their corresponding redshifts $z_i$ and $z_j$, we regularize the log-likelihood as
\begin{align}
D_{ij} &= \frac{1-\exp\left(-\frac{(z_i-z_j)^2}{2\rho_\mathrm{corr}^2}\right)}{2\pi\sqrt{\rho_\mathrm{corr}}} \\
\ln \mathcal L_{\mathrm{reg}ij} &= -\ln D_{ij}- \frac12 \left(\frac{A_i-A_j}{D_{ij}}\right)^2 + \mathrm{const.}
\end{align}
The total log-likelihood is the sum over all cluster likelihoods [see Eq.~(\ref{eq:boost_likelihood})] and over all regularization terms,
\begin{equation}
\ln \mathcal L = \sum_i \ln \mathcal L_{\mathrm{cluster},i} + \sum_j \ln \mathcal L_{\mathrm{reg},j}  + \mathrm{const.}
\end{equation}

The SOMPZ redshift estimates (see Sec.~\ref{sec:DES_Pz}) turn out to be inadequate to estimate the cluster member contamination and we are not able to extract a meaningful measurement.
Therefore, we estimate the cluster member contamination using the DNF and BPZ redshift estimates.
We explain the better performance with the fact that DNF and BPZ are trained on optimized photometry made in all DES bands, whereas the SOMPZ are restricted to the \textsc{Metacalibration} $r,i,z$ photometry.
In practice, we construct the DNF and BPZ source redshift distributions using each source's point estimate $z_\mathrm{mc}$ instead of its full redshift probability distribution.

\begin{table*}
\caption{\label{tab:clmemcont}
Parameters of the cluster member contamination model (mean and standard deviation).}
\begin{ruledtabular}
\begin{tabular}{lcccc}
Parameter & \multicolumn{2}{c}{DNF photo-$z$} & \multicolumn{2}{c}{BPZ photo-$z$}\\
 & SPT center & MCMF center & SPT center & MCMF center\\
\colrule
$z_\mathrm{off,0}$ & $0.0523\pm0.0018$ & $0.0513\pm0.0018$ & $0.0273\pm0.0010$ & $0.0269\pm0.0011$\\
$z_\mathrm{off,1}$ & $-0.1341\pm0.0105$ & $-0.1346\pm0.0111$ & $-0.1320\pm0.0077$ & $-0.1303\pm0.0075$\\
$\sigma_{z,0}$ & $0.0885\pm0.0016$ & $0.0877\pm0.0016$ & $0.0783\pm0.0011$ & $0.0784\pm0.0011$\\
$\sigma_{z,1}$ & $-0.0497\pm0.0086$ & $-0.0531\pm0.0095$ & $-0.0375\pm0.0069$ & $-0.0367\pm0.0068$\\
$\log(c)$ & $0.545\pm0.028$ & $0.504\pm0.027$ & $0.676\pm0.028$ & $0.652\pm0.028$\\
$B_\lambda$ & $0.703\pm0.028$ & $0.730\pm0.029$ & $0.750\pm0.024$ & $0.773\pm0.025$\\
$\rho_\mathrm{corr}$ & $0.1021\pm0.0038$ & $0.1018\pm0.0038$ & $0.1065\pm0.0041$ & $0.1048\pm0.0040$\\
$A_0$ & $-1.14\pm0.52$ & $-1.08\pm0.52$ & $-0.66\pm0.47$ & $-0.69\pm0.49$\\
$A_1$ & $0.26\pm0.45$ & $0.13\pm0.44$ & $0.47\pm0.40$ & $0.33\pm0.42$\\
$A_2$ & $1.23\pm0.36$ & $1.26\pm0.36$ & $1.24\pm0.32$ & $1.22\pm0.33$\\
$A_3$ & $-0.17\pm0.42$ & $-0.15\pm0.41$ & $0.06\pm0.37$ & $0.08\pm0.38$\\
$A_4$ & $-0.23\pm0.39$ & $-0.33\pm0.38$ & $-0.11\pm0.35$ & $-0.23\pm0.36$\\
$A_5$ & $1.25\pm0.39$ & $1.27\pm0.38$ & $1.26\pm0.34$ & $1.25\pm0.35$\\
$A_6$ & $0.87\pm0.43$ & $0.85\pm0.42$ & $1.02\pm0.39$ & $1.02\pm0.40$\\
$A_7$ & $-0.32\pm0.41$ & $-0.37\pm0.41$ & $-0.28\pm0.38$ & $-0.38\pm0.38$\\
$A_8$ & $-0.09\pm0.41$ & $-0.06\pm0.41$ & $-0.08\pm0.36$ & $-0.16\pm0.38$\\
$A_9$ & $0.85\pm0.47$ & $0.81\pm0.46$ & $0.77\pm0.43$ & $0.83\pm0.44$\\
$A_{10}$ & $1.21\pm0.57$ & $1.04\pm0.57$ & $0.86\pm0.54$ & $0.94\pm0.55$\\
$A_\infty$ & $-3.74\pm0.83$ & $-3.69\pm0.82$ & $-4.02\pm0.75$ & $-3.87\pm0.78$\\
\end{tabular}
\end{ruledtabular}
\end{table*}

\begin{figure}
  \includegraphics[width=\columnwidth]{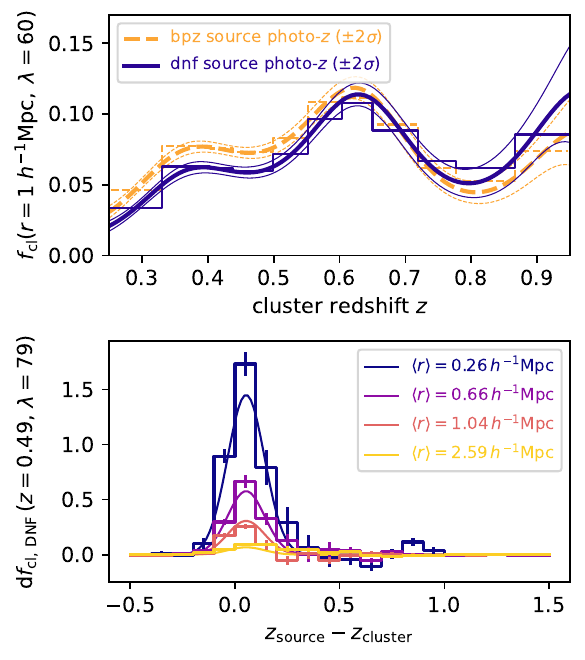}
  \caption{Fractional cluster member contamination $f_\mathrm{cl}$ of the weak-lensing shear signal.
  \emph{Upper panel:} Evolution as a function of cluster redshift, using the DNF and BPZ photo-$z$ estimators.
  Thick lines show the mean and thin lines show the 2-$\sigma$ interval.
  As a cross check, we also show the mean result obtained using bins in redshift instead of the smooth function.
  \emph{Lower panel:} Example redshift distributions of the contaminants for clusters with $0.4<z<0.6$ and $60<\lambda<100$.
  Thin, smooth lines show the model, which is also used in the top panel and throughout our weak-lensing analysis.
  Binned histograms show the distribution as obtained using a model-free approach for cross-check purposes.
  The contaminants are concentrated around the cluster redshift, and their number decreases with increasing distance from the cluster center.
  }
  \label{fig:clmemcont}
\end{figure}

We report the recovered parameters of our cluster member contamination model in Table~\ref{tab:clmemcont}.
We show the evolution with redshift in the top panel of Fig.~\ref{fig:clmemcont}.
It is clear that the trend follows no obvious functional form which motivates our complex modeling, as was discussed in \citep{paulus_thesis}.
For cross-check purposes, we also consider a simpler model in which we replace the term for the redshift evolution [second line in Eq.~(\ref{eq:boost_model})] with a simple amplitude in bins of redshift.
The recovered results from the simplified analysis are also shown in the top panel of Fig.~\ref{fig:clmemcont}, and confirm the complex evolution with redshift.
Further note that the choice of cluster center (SPT or optical) has a very minor impact on the amount of cluster member contamination. This is expected because we do not consider the innermost 500~$h^{-1}$kpc in this analysis.
Finally, and most interestingly, there is a statistically significant deviation between the cluster member contamination as inferred using DNF redshifts or BPZ redshifts (purple vs. orange lines in top panel of Fig.~\ref{fig:clmemcont}).

To further validate our analysis, we also perform a model-free measurement of cluster member contamination.
Still following \cite{paulus_thesis}, we note that the cluster member contamination is localized around the cluster redshift.
Therefore, we assume that there is no residual cluster member contamination at $z_\mathrm{cluster} + 0.5$.
We can then use the field redshift distribution, measured far away from the cluster center, and the cluster line-of-sight redshift distribution, and normalize both so that they match above $z_\mathrm{cluster}+0.5$.
Any local enhancement of the cluster line-of-sight redshift distribution can then be attributed to cluster members.
Using DNF and BPZ source redshifts, we apply this test to stacked measurements of all clusters within three bins in redshift and two bins in richness.
The bottom panel of Fig.~\ref{fig:clmemcont} shows one such analysis for clusters with $0.4<z<0.6$ and richness $60<\lambda<100$ using DNF redshifts.
In the radial range of interest, the model-free estimations and our recovered model agree reasonably well.
This test confirms that the BPZ redshifts do indeed indicate a higher level of cluster member contamination than the DNF redshifts.

Given the apparent mismatch between the cluster member contamination as determined from BPZ and DNF redshifts, we will perform a blind comparison of the cosmological results obtained from the real data using either model (\citetalias{bocquet24II}).
Another robustness test will consist in only using lensing data for $r>800~h^{-1}$kpc instead of $r>500~h^{-1}$kpc, thereby excluding more of the radial range where the cluster member contamination is particularly strong.

\section{The Cluster Weak-Lensing Model}
\label{sec:clusterlensmodel}

We reviewed the theory of cluster lensing in Sec.~\ref{sec:cluster_lens_basics} and we discussed the measurements of lensing shear profiles in Sec.~\ref{sec:DESmeasure}.
Here, we introduce the model we employ for the DES~Y3 lensing dataset and summarize the HST lensing model.

\subsection{The Model for HST Weak Lensing}
\label{sec:HSTmodel}

The model for the HST lensing data was introduced along with the measurements in \cite{schrabback18, schrabback21, zohren22} and implemented in \cite{dietrich19, bocquet19}.
Here, we briefly summarize the key points and refer the reader to the referenced works.
We model the HST shear profiles using the NFW profile and the halo-concentration--mass relation $c(M,z)$ from \cite{diemer&joyce19}, along with the measured source redshift distributions.
The residual mismatch between real, miscentered halo profiles (and their diversity) and the NFW model is captured in an \Mwl--\Mhalo\ relation, where \Mwl\ is the masslike quantity that enters the NFW model and \Mhalo\ is the halo mass definition we adopt in modeling the halo mass function.
More details of the mass modeling are also found in \cite{sommer22}.
The uncertainties in the source redshift distribution and shear calibration and the effects of
line-of-sight variations in the matter and source redshift distributions are quantified and accounted for in the analysis.

\subsection{The Model for DES Weak Lensing}
\label{sec:DESmodel}

We now describe the model we adopt for relating the DES weak-lensing measurements to the underlying halo mass.
The methodology is developed in \cite{grandis21} where a generic but realistic toy model is considered.
We compute the surface mass density profile starting from an NFW profile with a constant halo concentration $c=3.5$ and an approximate correction for miscentering of magnitude $R_\mathrm{mis}$ [see Sec.~\ref{sec:miscentering} and Eq.~(\ref{eq:Rmis})]:
\begin{equation} \label{eq:Sigma_of_r}
 \Sigma(r|M)= 
\begin{cases}
    \Sigma_\mathrm{NFW}(R_\mathrm{mis}|M, c)  &\text{for } r\leq R_\mathrm{mis} \\
    \Sigma_\mathrm{NFW}(r| M,c) &\text{for } r>R_\mathrm{mis}.
\end{cases}
\end{equation}
Note that, because $\Sigma(r|M)$ is constant within $R_\mathrm{mis}$, the density contrast vanishes for these radii [see Eq.~(\ref{eq:gamma})],
\begin{equation}
    \Delta\Sigma(r\leq R_\mathrm{mis}) = 0.
\end{equation}
The density contrast outside of $R_\mathrm{mis}$ is computed as
\begin{equation}
\label{eq:DeltaSigma_out}
  \begin{split}
    \Delta\Sigma(r>R_\mathrm{mis}) \equiv& \langle \Sigma(<r)\rangle - \Sigma(r) \\
    =& \langle \Sigma_\mathrm{NFW}(<r)\rangle + \frac{R_\mathrm{mis}^2}{r^2} \bigl[\Sigma_\mathrm{NFW}(R_\mathrm{mis}) \\
    &- \langle \Sigma_\mathrm{NFW}(<R_\mathrm{mis})\rangle\bigr] - \Sigma_\mathrm{NFW}(r) \\
    =& \Delta\Sigma_\mathrm{NFW}(r) - \frac{R_\mathrm{mis}^2}{r^2} \Delta\Sigma_\mathrm{NFW}(R_\mathrm{mis}).
  \end{split}
\end{equation}
Note that since $\Sigma_\mathrm{NFW}(r)$ and $\Delta\Sigma_\mathrm{NFW}(r)$ have analytical solutions (see, e.g., \cite{wright00}), Eq.~(\ref{eq:DeltaSigma_out}) can be computed exactly.
We now use $\Sigma(r)$ and $\Delta\Sigma(r)$ to compute the shear profile using Eqs.~(\ref{eq:gamma}) and (\ref{eq:gt}).
We compute $\langle\Sigma_\mathrm{crit}^{-1}\rangle$ [Eq.~(\ref{eq:Sigmacrit})] using the source redshift distribution [Eq.~(\ref{eq:source_redshift_distribution})].
Finally, we account for the mean effect of cluster member contamination by correcting the model shear profile with $1-f_\mathrm{cl}(r,z_\mathrm{cluster},\lambda)$ [see Eq.~(\ref{eq:f_cl})].
In summary, our model shear profile is constructed from an NFW profile, is approximately corrected for miscentering, and corrected for the mean amount of cluster member contamination.
Since the lensing efficiency $\langle\invSigmac\rangle$ explicitly depends on cosmology, we re-compute it at each step in the likelihood analysis following Eq.~(\ref{eq:meaninvSigmac}).

We follow the discussion in \cite{grandis21} and only consider the radial range between 500~$h^{-1}$kpc and $3.2/(1+z_\mathrm{cluster})~h^{-1}$Mpc.
The inner limit avoids the regime where miscentering, cluster member contamination, and hydrodynamical effects play a more significant role.
The outer limit is chosen to exclude the 2-halo term regime; our lensing analysis is thus restricted to the 1-halo-term regime.
As a cross-check of our fiducial analysis choice, in \citetalias{bocquet24II}, we will also perform an analysis where we exclude the innermost 800~$h^{-1}$kpc.

\subsubsection{Weak-Lensing Mass Bias and Scatter}
\label{sec:massbiasscatter}

Our simple model for the shear profile is not a perfect description of actual shear profiles. In particular, it does not account for departures from the NFW profile or for all sources of uncertainty.
Therefore, the halo mass inferred using this model is a biased and noisy mass estimator.
Rather than making the model more complex, we define the mass $M$ that enters Eq.~(\ref{eq:Sigma_of_r}) as a latent variable called the ``weak-lensing mass'' $M_\mathrm{WL}$, and establish an \Mwl--\Mhalo\ relation, where \Mhalo\ is mass for which the halo mass function is defined.
We calibrate this relationship accounting for the fact that we model the complex halo projected mass distributions with a simplistic model (this is also referred to as halo mass modeling) and accounting for the observational systematic and stochastic uncertainties.\footnote{Note that alternatively, one could explicitly marginalize over the sources of lensing uncertainty during the cosmological likelihood analysis. We tested such an approach and concluded that it is computationally intractable for the size and complexity of the weak-lensing dataset considered here.}

We follow the methodology presented in \cite{grandis21} to calibrate the \Mwl--\Mhalo\ relation.
From the \textit{Magneticum} simulation suite \citep{hirschmann14, teklu15, beck16, dolag17}, we use pairs of hydrodynamical and gravity-only runs with identical initial conditions to create the link between the gravity-only halo mass and realistic, full-physics halo mass profiles.
This allows us to use accurate predictions for the halo mass function from gravity-only simulations (\cite{tinker08}, but also emulators \citep{mcclintock19emu, nishimichi19, bocquet20}, while simultaneously accounting for the impact of baryonic effects on halo profiles and thus on cluster cosmology (i.e., we argue that our approach addresses the concerns raised in, e.g., \cite{debackere21}).
We then repeat the same analysis but use the Illustris-TNG hydrodynamical simulations \citep{pillepich18, marinacci18, springel18, nelson18, naiman18, nelson19}.
Finally, we estimate the impact of the uncertainty in baryonic effects on the \Mwl--\Mhalo\ relation by taking the difference of the results based on \textit{Magneticum} and Illustris-TNG.\footnote{While estimating an uncertainty by comparing two sets of results is not ideal, it reflects the status quo. In future work, we will compare the calibrations obtained from more numerical simulations as they become available.}

The strategy for calibrating the $\Mwl$--$\Mhalo$ relation is to use the projected mass maps from numerical simulations to create synthetic lensing shear profiles according to the specifications of the DES~Y3 lensing measurements of SPT clusters. We now summarize these specifications and their implementation:
\begin{enumerate}
    \item For each halo in the simulation with $M_{200\mathrm{c}}>1.56\times10^{14}~h^{-1}\Msun$, we create three sets of two-dimensional mass maps by projecting along the three orthogonal directions, with a projection depth of $\pm\,20\,h^{-1}$Mpc.
    In practice, we down-sample the more abundant low-mass halo population to achieve a roughly constant number of halos per logarithmic mass interval.
    We analyze 9,798~mass maps for a total of 3,266~halos;

    \item For each halo mass map, we define a set of positions that are offset from the true halo center by an array of radii $R_\text{mis}$; the azimuthal angle is drawn uniformly.\footnote{Drawing the direction of miscentering uniformly neglects the potential correlation between miscentering and halo morphology, which can bias the inferred lensing mass \citep{sommer23}. In the cosmological analysis of the real dataset, we will compare the results obtained using optical and SPT centers, or a large radial cut (and the corresponding models) as cross-checks.}
    We then process the projected mass maps into polarly binned, scaled maps of convergence $\Sigma (R,\,\phi | R_\text{mis})$ and tangential shear $ \Gamma_\mathrm{t}(R,\,\phi | R_\text{mis})$ for each set of polar positions;

    \item We construct synthetic source redshift distribution $P^\text{synth}(z_\text{s} )$ as in Eq.~(\ref{eq:source_redshift_distribution}), but we randomly draw the distributions $P_b(z_\mathrm{s})$ (that include the multiplicative shear bias $m$) from the 1,000 realizations of the calibration systematics, to capture the impact of these systematic uncertainties;

    \item Using the maps of convergence and shear and the source redshift distributions, we now produce synthetic tangential shear profiles.
    Improving upon previous work, we compute the reduced shear not only for each polar position in the map, but also for each source redshift $z_\text{s}$.
    Averaging over azimuth and source redshifts is done after the computation of the reduced shear, and the mean profile is
    \begin{equation}
    \begin{split}
        g_t^\text{synth}(R| R_\text{mis}) = \int& \text{d} z_\text{s}~ P^\text{synth}(z_\text{s} ) \int \frac{\text{d}\phi}{2 \pi} \\
        &\frac{\Sigma_\text{crit,ls}^{-1} ~\Gamma_t(R,\,\phi | R_\text{mis})}{1-\Sigma_\text{crit,ls}^{-1}~ \Sigma (R,\,\phi | R_\text{mis}) }.
    \end{split}
   \end{equation}
   Deviating from this order in the integration would bias the synthetic profiles at the level of 0.01, which would not be acceptable given our targeted level of accuracy.
   Note that low-order corrections for this bias exist \citep{seitz97}.
   However, instead of complicating the model with such a correction, we prefer to absorb the bias into a correct model of the synthetic shear profiles and thus into the lensing bias we are in the process of calibrating;
   
    \item Our models for miscentering and for cluster member contamination depend on the cluster richness. Therefore, for each halo in the simulation, we draw a richness according to the scaling relation in Eq.~(\ref{eq:lambdaM}), with scatter given by a combination of the intrinsic lognormal scatter $\sigma_{\ln\tilde\lambda}$ and a Poisson contribution.
    The parameters of the richness--mass relation are drawn as
    \begin{equation}
      \begin{split}
        A_\lambda &\sim\mathcal N(76.5, 8.2^2), \\
        B_\lambda &\sim\mathcal N(1.02, 0.08^2), \\
        C_\lambda &\sim\mathcal N(0.29, 0.27^2), \\
        \sigma_{\ln\tilde\lambda} &\sim\mathcal N\left(\ln 0.23, (0.16/0.23)^2\right),
      \end{split}
    \end{equation}
    as given in \cite[Table 4]{bleem20};
    
    \item We apply the effects of the shape measurement bias and cluster member contamination to the synthetic shear profiles $g_t^\text{synth}(R)$, accounting also for possible non-linear shear biases;\footnote{The non-linear shear bias $\alpha_\mathrm{NL}$ incorporates, among others, the potential biases arising when measuring shapes in crowded cluster fields. We thus marginalize over a generous prior $\ln\alpha_\mathrm{NL}\sim \mathcal N(\ln~0.6, 0.4^2)$ \citep[following][]{sheldon&huff17, mcclintock19wl, grandis21}.}

    \item We draw off-centered cluster positions from the calibrated miscentering distributions.
    Note that we account for the stochastic noise and the systematic uncertainty in the miscentering model.
    To draw from the SPT miscentering distribution, we first assign core radii $\theta_c$ and detection significances $\xi$ to the simulated halos [see Eq.~(\ref{eq:SPT_pos_unc})].
    The distribution of core radii is well-described by an exponential distribution
    \begin{equation}
      \theta_c\, D_A(z) \sim R_{c,0} \exp\left( - \frac{\theta_c\, D_A(z)}{R_{c,0}}\right),
    \end{equation}
    with the angular diameter distance $D_A(z)$.
    We determine the scale
    \begin{equation}
        R_{c,0}^{-1} = 3.76 \pm 0.16\,h/\mathrm{Mpc}
    \end{equation}
    which we adopt as a prior, assuming no variation with mass or redshift. To predict $\xi$, which modulates the strength of the observational positional uncertainty, we follow the scaling relation and scatter model described in Eqs.~(\ref{eq:xizeta})--(\ref{eq:aszfield}), with priors on the SZ scaling relation parameters
    \begin{equation}
      \begin{split}
        A_\mathrm{SZ} &\sim\mathcal N(5.24, 0.85^2), \\
        B_\mathrm{SZ} &\sim\mathcal N(1.53, 0.1^2), \\
        C_\mathrm{SZ} &\sim\mathcal N(0.47, 0.41^2), \\
        \sigma_{\ln\zeta} &\sim\mathcal N\left(\ln 0.27, (0.1/0.27)^2\right),
      \end{split}
    \end{equation}
    as given in \cite{bocquet19}.
    We furthermore assume a field depth of $\gamma_\text{field}=1.2$ and draw $\xi$ from a truncated Gaussian, such that $\xi>4.5$ to avoid divisions by very small values of $\xi$ in Eq.~(\ref{eq:SPT_pos_unc});
    \item We create realizations of the cluster member contamination model;
    
    \item We compute realizations of shear due to the uncorrelated large-scale structure along the line of sight.
\end{enumerate}

\begin{table*}
\scriptsize
\caption{\label{tab:WLmodel}
Parameters of the weak-lensing-mass-to-halo-mass (\Mwl--\Mhalo) relation (mean and standard deviation). In the cosmological analysis, we use the full posterior distribution to also account for the parameter covariances. The redshifts are $z\in\{0.252, 0.470, 0.783, 0.963\}$.}
\begin{ruledtabular}
\begin{tabular}{lcccccccc}
Parameter & \multicolumn{4}{c}{Optical center} & \multicolumn{4}{c}{SPT center}\\
 & \multicolumn{2}{c}{$r>500~h^{-1}$kpc} & \multicolumn{2}{c}{$r>800~h^{-1}$kpc} & \multicolumn{2}{c}{$r>500~h^{-1}$kpc} & \multicolumn{2}{c}{$r>800~h^{-1}$kpc}\\
 & DNF & BPZ & DNF & BPZ & DNF & BPZ & DNF & BPZ\\
\colrule
$\ln b_\mathrm{WL}(z_0)$ & $-0.042$ & $-0.044$ & $-0.055$ & $-0.056$ & $-0.007$ & $-0.009$ & $-0.022$ & $-0.022$ \\
$\ln b_\mathrm{WL}(z_1)$ & $-0.040$ & $-0.046$ & $-0.058$ & $-0.061$ & $0.005$ & $-0.002$ & $-0.017$ & $-0.018$ \\
$\ln b_\mathrm{WL}(z_2)$ & $-0.033$ & $-0.038$ & $-0.083$ & $-0.075$ & $0.025$ & $0.024$ & $-0.018$ & $-0.014$ \\
$\ln b_\mathrm{WL}(z_3)$ & $-0.082$ & $-0.089$ & $-0.163$ & $-0.145$ & $-0.015$ & $-0.012$ & $-0.088$ & $-0.074$ \\
$\sigma_{\ln b_\mathrm{WL,1}}(z_0)$ & $-0.006$ & $-0.005$ & $-0.005$ & $-0.005$ & $-0.006$ & $-0.006$ & $-0.005$ & $-0.005$ \\
$\sigma_{\ln b_\mathrm{WL,1}}(z_1)$ & $-0.014$ & $-0.013$ & $-0.013$ & $-0.012$ & $-0.013$ & $-0.013$ & $-0.013$ & $-0.011$ \\
$\sigma_{\ln b_\mathrm{WL,1}}(z_2)$ & $-0.052$ & $-0.054$ & $-0.055$ & $-0.051$ & $-0.053$ & $-0.052$ & $-0.052$ & $-0.051$ \\
$\sigma_{\ln b_\mathrm{WL,1}}(z_3)$ & $-0.112$ & $-0.114$ & $-0.115$ & $-0.105$ & $-0.120$ & $-0.110$ & $-0.110$ & $-0.104$ \\
$\sigma_{\ln b_\mathrm{WL,2}}(z_0)$ & $0.008$ & $0.008$ & $0.008$ & $0.007$ & $-0.009$ & $0.010$ & $0.007$ & $0.008$ \\
$\sigma_{\ln b_\mathrm{WL,2}}(z_1)$ & $0.015$ & $0.016$ & $0.013$ & $0.015$ & $-0.015$ & $0.016$ & $0.013$ & $0.014$ \\
$\sigma_{\ln b_\mathrm{WL,2}}(z_2)$ & $0.017$ & $0.014$ & $0.015$ & $0.013$ & $-0.019$ & $0.014$ & $0.016$ & $0.012$ \\
$\sigma_{\ln b_\mathrm{WL,2}}(z_3)$ & $-0.010$ & $-0.009$ & $-0.009$ & $-0.008$ & $0.010$ & $-0.009$ & $-0.009$ & $-0.008$ \\
$b_{\mathrm{WL},M}$ & $1.029\pm0.006$ & $1.027\pm0.007$ & $1.049\pm0.007$ & $1.047\pm0.007$ & $0.995\pm0.008$ & $0.993\pm0.008$ & $1.017\pm0.007$ & $1.015\pm0.007$ \\
$s_\mathrm{WL}(z_0)$ & $-3.115\pm0.044$ & $-3.112\pm0.042$ & $-2.892\pm0.034$ & $-2.888\pm0.034$ & $-3.040\pm0.047$ & $-3.038\pm0.049$ & $-2.872\pm0.034$ & $-2.871\pm0.034$ \\
$s_\mathrm{WL}(z_1)$ & $-3.074\pm0.048$ & $-3.071\pm0.046$ & $-2.840\pm0.035$ & $-2.833\pm0.034$ & $-2.980\pm0.056$ & $-2.976\pm0.054$ & $-2.817\pm0.035$ & $-2.814\pm0.034$ \\
$s_\mathrm{WL}(z_2)$ & $-2.846\pm0.060$ & $-2.847\pm0.057$ & $-2.427\pm0.048$ & $-2.429\pm0.047$ & $-2.711\pm0.072$ & $-2.709\pm0.074$ & $-2.418\pm0.047$ & $-2.421\pm0.048$ \\
$s_\mathrm{WL}(z_3)$ & $-1.945\pm0.101$ & $-1.952\pm0.104$ & $-1.378\pm0.086$ & $-1.393\pm0.084$ & $-1.842\pm0.102$ & $-1.877\pm0.101$ & $-1.370\pm0.089$ & $-1.383\pm0.085$ \\
$s_{\mathrm{WL},M}$ & $-0.226\pm0.040$ & $-0.239\pm0.041$ & $-0.590\pm0.043$ & $-0.595\pm0.043$ & $-0.302\pm0.043$ & $-0.315\pm0.044$ & $-0.601\pm0.043$ & $-0.606\pm0.043$
\end{tabular}
\end{ruledtabular}
\end{table*}

We now have libraries of synthetic shear profiles, which are created accounting for all relevant sources of statistical and systematic uncertainties. Each halo in this synthetic catalog has a shear profile for each projection direction and miscentering distance. Each of these shear profiles are then fitted with our extraction model described above, resulting in a weak-lensing mass \Mwl.
Given the known halo mass in the simulation, we can now calibrate the \Mwl--\Mhalo\ relation by weighting the individual weak-lensing masses with the miscentering distribution (see Section 2.3.1 in Ref.~\cite{grandis21}).
The mean relation is well-described by 
\begin{equation}
    \left\langle\ln \left(\frac{M_\mathrm{WL}}{M_0}\right)\right\rangle = b_\mathrm{WL}(z) + b_{\mathrm{WL},M} \ln\left(\frac{M_{200\mathrm{c}}}{M_0}\right) 
    \label{eq:MwlM}
\end{equation}
with a pivot mass $M_0 = 2\times10^{14}~h^{-1}\Msun$.
The scatter in $\ln M_\mathrm{WL}$ is well-described by a normal distribution of width
\begin{equation}
    \ln\sigma_{\ln M_\mathrm{WL}} = \frac12 \left[s_\mathrm{WL}(z) + s_{\mathrm{WL},M} \ln\left(\frac{M_{200\mathrm{c}}}{M_0}\right)\right]
    \label{eq:sigmalnMwl}
\end{equation}
with the same value for $M_0$.
In practice, we use simulations at four redshifts $z\in\{0.252, 0.470, 0.783, 0.963\}$ to calibrate the free parameters of the model.
To correctly capture the somewhat complex dependence of the uncertainty on $b_\mathrm{WL}$ with redshift, we describe $\sigma_{b_\mathrm{WL}}(z)$ as the linear combination of two independent components, determined via principal component analysis,
\begin{equation}
    \sigma_{b_\mathrm{WL}}(z) = \sigma_{b_\mathrm{WL,1}}(z) + \sigma_{b_\mathrm{WL,2}}(z).
\end{equation}
To obtain values for the bias or scatter at any intermediate redshift, we interpolate linearly.

\begin{figure}
  \includegraphics[width=\columnwidth]{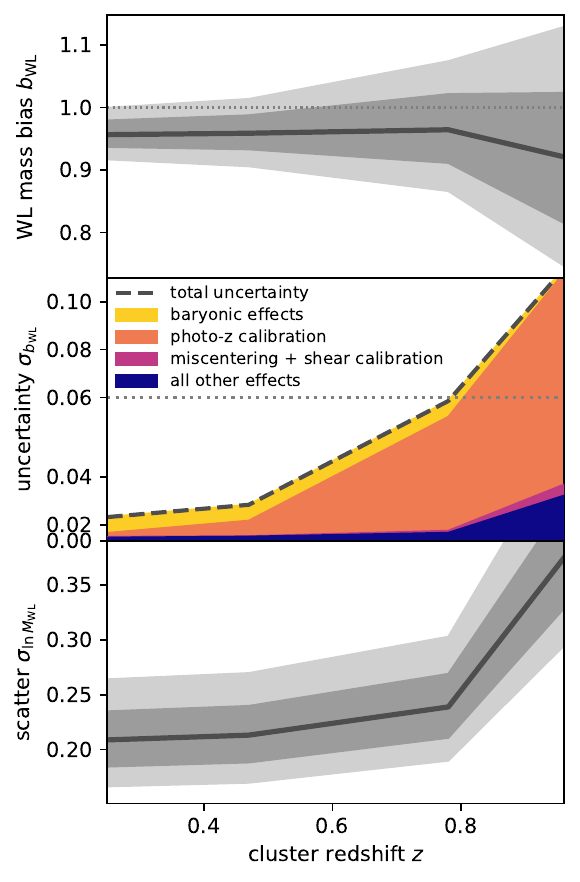}
  \caption{Evolution of the lensing mass bias $b_\mathrm{WL}$, its uncertainty $\sigma_{b_\mathrm{WL}}$, and the intrinsic scatter \sigmawl\ with redshift, at the pivot mass $2\times10^{14}~h^{-1}\Msun$.
  \emph{Top and bottom panels:} Solid lines show the mean relations, and the shaded bands show the 68\% and 95\% credible intervals.
  \emph{Middle panel:}
  Contribution of the error budgets of individual model components to the total uncertainty in $b_\mathrm{WL}$ of about $2-10\%$.
  The error budget due to ``all other effects'' contains, among others, the impact of the uncertainty in the cluster member contamination and of the noise due to the finite set of simulated halos.
  The total uncertainty is largely dominated by the uncertainties in the source photo-$z$ calibration and the impact of baryonic effects.}
  \label{fig:lensingbiasscatter}
\end{figure}

We compute eight sets of weak-lensing bias and scatter posteriors, varying the centers (SPT vs optical), photo-z codes used for the estimation of the cluster member contamination (BPZ vs DNF), and the inner fitting radius ($R_\text{min}=0.5,\,0.8~h^{-1}$Mpc).
The bias and scatter parameters are summarized in Table~\ref{tab:WLmodel}.
As discussed, this model is established based on the \textit{Magneticum} simulations.
We repeat the same analysis using two snapshots of Illustris-TNG, or 1,431~mass maps from 477~halos.
The recovered model parameters differ from the ones based on \textit{Magneticum} as follows (see also Sec.~3.4 in Ref.~\cite{grandis21}):
\begin{equation}
  \begin{split}
    \Delta b_\mathrm{WL} = 0.02, \\
    \Delta b_{\mathrm{WL},M} = 0.018, \\
    \Delta s_\mathrm{WL} = 0.25, \\
    \Delta s_{\mathrm{WL},M} = 0.59.
  \end{split}
\end{equation}
We interpret these differences as uncertainties in the \Mwl--\Mhalo\ relation due to baryonic effects, and add them in quadrature to the uncertainties quoted in Table~\ref{tab:WLmodel}.
In the likelihood analysis, we sample the bias and scatter parameters within these combined uncertainties.
The top and bottom panels in Fig.~\ref{fig:lensingbiasscatter} show the evolution of the mass bias and scatter with cluster redshift.
In the analysis of the real data in \citetalias{bocquet24II}, we will show that the parameter uncertainties in the lensing model are subdominant in comparison to the measurement errors.
This justifies our approach of estimating the impact of the uncertainties in baryonic modeling using only two sets of simulations.
For future work, we plan to compare more simulations to obtain a more refined error estimate.

\subsubsection{Discussion of the Lensing Mass Bias and Scatter}
\label{sec:massbiasscatterdiscussion}

We now discuss the impact of the various elements that enter the determination of the lensing bias and scatter (see bullet list in the previous subsection).
We do so by examining the (squared) correlation coefficients $\rho_{i,j}^2$ between the effect $i$ under consideration (e.g., photo-$z$ calibration) and the output quantity $j$ of interest (e.g., uncertainty on lensing mass bias $b_\mathrm{WL}$) (see Sec.~2.3.4 in Ref.~\cite{grandis21}).

We illustrate the impact of a selection of effects on the final uncertainty in the weak-lensing mass bias in the middle panel of Fig.~\ref{fig:lensingbiasscatter}.
The final uncertainty is dominated by the uncertainties in baryonic effects at low redshifts and the photo-$z$ calibration (through $\langle \Sigma_\text{crit}^{-1}\rangle$) at cluster redshifts beyond around 0.45.
The uncertainty due to the combined effects of the finite size of the simulated halo sample, miscentering, cluster member contamination, and shear calibration is small and it only amounts to $\sim1\%$ uncertainty up to cluster redshift $z\sim0.8$.

The uncertainty on the weak-lensing scatter correlates less strongly with the individual model components.
This suggests that the limitation of having a relatively small sample of simulated halos is more important here than it is for determining the bias.

The \Mwl--\Mhalo\ calibration requires us to assume a fiducial richness--mass and SZ--mass relation (see items 5 and 7 in Sec.~\ref{sec:massbiasscatter}).
Given that both the synthetic shear profiles and our model have very similar dependencies on richness via miscentering and cluster member contamination, the width of the priors on the richness--mass relation parameters do not affect the uncertainty on the weak-lensing bias and scatter (the squared correlation coefficient is small).
The situation is analogous for the parameters of the SZ--mass relation, which enters through the SPT positional uncertainty.
In summary, the choice of fiducial observable--mass relations is necessary to calibrate our lensing model but it does not affect our final result strongly.

As discussed in Sec.~\ref{sec:miscenter_sims}, our data-driven SZ miscentering distribution does not agree well with that extracted from the \textit{Magneticum} simulations, which could point to some limitations of the simulation data products.
In this section, we use these same simulations (and Illustris TNG) to calibrate the \Mwl--\Mhalo\ relation.
In our lensing analysis, we only use scales beyond $500~h^{-1}$kpc that are much larger than the typical offset in SPT centers (see Fig.~\ref{fig:miscenter}).
Therefore, we expect our analysis to be robust to the shortcomings of simulating the challenging cluster central regions.

\section{SZ and Richness Scaling Relations}
\label{sec:OMR}

As in previous SPT work, the SZ detection significance $\xi$ is related to the unbiased significance $\zeta$ (e.g.,~\cite{vanderlinde10})
\begin{equation} \label{eq:xizeta}
P(\xi|\zeta) = \mathcal N\left(\sqrt{\zeta^2 + 3}, 1\right).
\end{equation}
This relationship accounts for the maximization bias in $\xi$ with respect to three free parameters (R.A., Dec., and filter scale) and the unit noise in the appropriately rescaled maps.
We assume lognormal intrinsic scatter in $\zeta$ of width $\sigma_{\ln\zeta}$.
The mean unbiased significance is modeled as a power-law relation in mass and $E(z)\equiv H(z)/H_0$
\begin{equation}
  \begin{split}
    \langle\ln\zeta\rangle =& \ln A_\mathrm{SZ} + B_\mathrm{SZ} \ln\left(\frac{M_{200\mathrm{c}}}{3\times 10^{14}~h^{-1}\Msun}\right) \\
    & + C_\mathrm{SZ} \ln\left(\frac{E(z)}{E(0.6)}\right). \label{eq:zetaM}
  \end{split}
\end{equation}
To account for the variable depth of the SPT surveys and fields, we rescale $A_\mathrm{SZ}$ for each individual SPT field
\begin{equation}
  \label{eq:aszfield}
  A_\mathrm{SZ,field} = \gamma_\mathrm{field} A_\mathrm{SZ}.
\end{equation}
The variations in depth also affect the redshift evolution $C_\mathrm{SZ}$.
Within the SPT-SZ and SPTpol~ECS surveys, the variations of $C_\mathrm{SZ}$ across fields  are neglible \citep{bleem15, bleem20}.
Following \cite{bleem23}, we rescale $C_\mathrm{SZ}$ for each SPT survey, assuming the SPT-SZ survey as the reference:
\begin{equation}
  \begin{split}
    C_\text{SZ,~SPT-SZ} &= C_\mathrm{SZ}, \\
    C_\text{SZ,~SPTpol~ECS} &= C_\mathrm{SZ}-0.09, \\
    C_\text{SZ,~SPTpol~500d} &= C_\mathrm{SZ}+0.26.
  \end{split}
\end{equation}

Similarly, we model the mean relation between the intrinsic richness $\tilde\lambda$ and mass as a power law in mass and $(1+\text{redshift})$ (e.g., \cite{DESY1cl}),
\begin{equation}
  \begin{split}
    \left\langle\ln\tilde\lambda\right\rangle =& \ln A_\lambda + B_\lambda \ln\left(\frac{M_{200\mathrm{c}}}{3\times 10^{14}~h^{-1}\Msun}\right) \\
    &+ C_\lambda \ln\left(\frac{1+z}{1.6}\right). \label{eq:lambdaM}
  \end{split}
\end{equation}
We assume lognormal intrinsic scatter in $\tilde\lambda$ of width $\sigma_{\ln\tilde\lambda}$.
We model the observational error on the measured richness $\lambda$ as an additional lognormal distribution with a width that corresponds to the Poisson uncertainty (e.g., \cite{saro15}), such that
\begin{equation}
  \label{eq:P_lambda_tildelambda}
  P\left(\ln\lambda|\ln\tilde\lambda\right) = \mathcal N\left( \ln\tilde\lambda, 1/\tilde\lambda\right).
\end{equation}

\section{Cluster Population Model}
\label{sec:likelihood}

Our analysis pipeline builds upon previous work, especially \cite{bocquet15, bocquet19}.
Significant updates have been required to handle the large amount of DES weak-lensing data.
We maintain the cluster-by-cluster weak-lensing mass calibration approach from previous analyses (as opposed to a stacking approach).

\subsection{Multi-Observable Scaling Relation}

The mean scaling relations between the unbiased SZ significance $\zeta$, optical richness $\tilde\lambda$, and weak-lensing mass \Mwl\ were defined in Eqs.~(\ref{eq:MwlM}), (\ref{eq:zetaM}), and (\ref{eq:lambdaM}).
As discussed, we model the intrinsic scatter in all observables as lognormal.
We account for possible correlations among all pairs of intrinsic scatter and establish a covariance matrix,
\begin{widetext}
\begin{equation} \label{eq:covmat}
    \vec\Sigma_\text{multi-obs} =
    \begin{pmatrix}
      \sigma_{\ln\zeta}^2 & \rho_\mathrm{SZ,WL}\sigma_{\ln\zeta}\sigmawl & \rho_\mathrm{SZ,\tilde\lambda}\sigma_{\ln\zeta}\sigma_{\ln\tilde\lambda} \\
      \rho_\mathrm{SZ,WL}\sigma_{\ln\zeta}\sigmawl & \sigmawl^2 & \rho_\mathrm{WL,\tilde\lambda}\sigmawl\sigma_{\ln\tilde\lambda} \\
      \rho_\mathrm{SZ,\tilde\lambda}\sigma_{\ln\zeta}\sigma_{\ln\tilde\lambda} & \rho_\mathrm{WL,\tilde\lambda}\sigmawl\sigma_{\ln\tilde\lambda} & \sigma_{\ln\tilde\lambda}^2
    \end{pmatrix}.
\end{equation}
We can now write the joint multi-observable scaling relation as a multivariate Gaussian distribution in log-observables
\begin{equation}
  \label{eq:scaling_relation}
    P\Bigl(
    \begin{bmatrix}
      \ln\zeta \\ \ln M_\mathrm{WL} \\ \ln\tilde\lambda
    \end{bmatrix} \big| M,z,\vec p\Bigr) = 
    \mathcal N\Bigl(
    \begin{bmatrix}
      \langle\ln\zeta\rangle(M,z,\vec p) \\ \langle\ln M_\mathrm{WL}\rangle(M,z,\vec p) \\ \langle \ln\tilde\lambda\rangle(M,z,\vec p)
    \end{bmatrix}, \vec\Sigma_\text{multi-obs}\Bigr)
\end{equation}
with the model parameters $\vec p$.

\subsection{Likelihood Function}

Neglecting sample variance (see Appendix~\ref{sec:samplevariance}), we describe the cluster population as (independent) Poisson realizations of the halo mass function.

\subsubsection{Poisson Likelihood}
The Poisson probability of observing $k$ events (halos) given the expected rate $\mu$ is
\begin{equation}
  P(k|\mu) = \frac{\mu^k e^{-\mu}}{k!} \Rightarrow \ln P(k|\mu) = k \ln\mu -\mu + \mathrm{const.}
\end{equation}
Splitting up our observable space in fine bins (in redshift, SPT detection significance, etc., such that each bin contains at most one event) we have a likelihood function,
\begin{equation}
  \label{eq:lnl_Poisson}
  \ln \mathcal L = \sum_i \ln\mu_i - \sum_j \mu_j,
\end{equation}
where the sum $i$ runs over all bins that contain an observed event, and the sum $j$ runs over all bins.\footnote{Of course one can also choose broader bins that contain more than one event; in this case, the second term in Eq.~(\ref{eq:lnl_Poisson}) needs to be scaled accordingly.}
We now take the limit of infinitesimally small bins $\dif x$.
The expected (differential) number of events then is $\dif\mu = \frac{\dif\mu}{\dif x}~\dif x$, and so
\begin{equation}
  \label{eq:poisson_dif}
    \ln \mathcal L = \sum_i \ln \left(\frac{\dif\mu}{\dif x}~\dif x\right)\Big|_{x_i} - \int \frac{\dif\mu}{\dif x}~\dif x = \sum_i \ln \frac{\dif\mu}{\dif x}\Big|_{x_i} - \int \frac{\dif\mu}{\dif x}~\dif x + \mathrm{const.}
\end{equation}
Note that in this form, the index $i$ runs over events, whereas above it ran over bins.
Therefore, in its differential form Eq.~(\ref{eq:poisson_dif}), the unbinned Poisson likelihood does indeed not involve any form of binning.

\subsubsection{Hierarchical Cluster Population Likelihood Function}

We now apply the Poisson likelihood to our multi-observable cluster sample:
\begin{equation}
    \label{eq:likelihood_start}
    \ln \mathcal L(\vec p) =  \sum_i \ln\frac{\dif^4 N(\vec p)}{\mathop{\dif\xi} \mathop{\dif\lambda} \mathop{\dif \vec g_\mathrm{t}} \mathop{\dif z}}
    \Big|_{\xi_i, \lambda_i, g_{\mathrm{t},i}, z_i}
     - \idotsint \mathop{\dif\xi} \mathop{\dif\lambda} \mathop{\dif \vec g_\mathrm{t}} \mathop{\dif z} 
     \frac{\dif^4 N(\vec p)}{\mathop{\dif\xi} \mathop{\dif\lambda} \mathop{\dif \vec g_\mathrm{t}} \mathop{\dif z}} \Theta_\mathrm{s}(\xi,\lambda,z)
     +\mathrm{const.}
\end{equation}
with the survey selection function $\Theta_\mathrm{s}$ which, in our analysis, is defined in terms of cuts in $\xi$, $\lambda$, and $z$ [see Eqs.~(\ref{eq:select_noDES}) and~(\ref{eq:select_DES})].
The lensing data are tangential shear profiles $\vec g_\mathrm{t}$.
The differential cluster abundance is
\begin{equation} \label{eq:d4N_dobs}
  \frac{\dif^4 N(\vec p)}{\mathop{\dif\xi} \mathop{\dif\lambda} \mathop{\dif \vec g_\mathrm{t}} \mathop{\dif z} } =
  \idotsint \mathop{\dif\Omega_\mathrm{s}}\mathop{\dif M} \mathop{\dif\zeta} \mathop{\dif\tilde\lambda} \mathop{\dif M_\mathrm{WL}}
  P(\xi|\zeta)
  P(\lambda|\tilde\lambda)
  P(\vec g_\mathrm{t}|M_\mathrm{WL}, \vec p)
  P(\zeta, \tilde\lambda, M_\mathrm{WL} |M,z,\vec p)
  \frac{\dif^2 N(M, z, \vec p)}{\mathop{\dif M} \mathop{\dif V}} \frac{\dif^2 V(z,\vec p)}{\mathop{\dif z} \mathop{\dif\Omega_\mathrm{s}}}
\end{equation}
with the halo mass function $\frac{\dif^2 N(M, z, \boldsymbol{p})}{\mathop{\dif M} \mathop{\dif V}}$ and the differential volume $\frac{\dif^2 V(z,\boldsymbol p)}{\mathop{\dif z} \mathop{\dif\Omega_\mathrm{s}}}$ within the survey footprint $\Omega_\mathrm{s}$.

The survey selection function $\Theta_\mathrm{s}$ is \emph{not} a function of lensing $\vec g_\mathrm{t}$, and so the second term in Eq.~(\ref{eq:likelihood_start}) becomes
\begin{equation}
    \iiiint \mathop{\dif\xi} \mathop{\dif\lambda} \mathop{\dif \vec g_\mathrm{t}} \mathop{\dif z} 
     \frac{\dif^4 N(\vec p)}{\mathop{\dif\xi} \mathop{\dif\lambda} \mathop{\dif \vec g_\mathrm{t}} \mathop{\dif z}} \Theta_\mathrm{s}(\xi,\lambda,z)
     =
     \int_{z_\mathrm{cut}}^\infty \dif z \int_{\xi_\mathrm{cut}}^\infty \dif\xi \int_{\lambda_\mathrm{min}(z)}^\infty \dif\lambda
     \frac{\dif^3 N(\vec p)}{\mathop{\dif\xi} \mathop{\dif\lambda} \mathop{\dif z}}
\end{equation}
with
\begin{equation}
    \frac{\dif^3 N(\vec p)}{\mathop{\dif\xi} \mathop{\dif\lambda} \mathop{\dif z} } =\int \mathop{\dif \vec g_\mathrm{t}} \frac{\dif^4 N(\vec p)}{\mathop{\dif\xi} \mathop{\dif\lambda} \mathop{\dif \vec g_\mathrm{t}} \mathop{\dif z} }=
    \iiiint \mathop{\dif\Omega_\mathrm{s}} \mathop{\dif M} \mathop{\dif\zeta} \mathop{\dif\tilde\lambda} P(\xi|\zeta) P(\lambda|\tilde\lambda)
 P(\zeta, \lambda |M,z,\vec p)
    \frac{\dif^2 N(M, z, \vec p)}{\mathop{\dif M} \mathop{\dif V}} \frac{\dif^2 V(z,\vec p)}{\mathop{\dif z} \mathop{\dif\Omega_\mathrm{s}}}.
\end{equation}
We can thus rewrite the log-likelihood from Eq.~(\ref{eq:likelihood_start}) and obtain our final log-likelihood,
\begin{equation}
  \label{eq:likelihood}
    \ln \mathcal L(\vec p) = \sum_i \ln \int_{\lambda_\mathrm{min}(z)}^\infty \mathop{\dif \lambda }\frac{\dif^3 N(\vec p)}{\mathop{\dif\xi} \mathop{\dif\lambda} \mathop{\dif z}} \Big|_{\xi_i, z_i}
    - \int_{z_\mathrm{cut}}^\infty \dif z \int_{\xi_\mathrm{cut}}^\infty \dif\xi \int_{\lambda_\mathrm{min}(z)}^\infty \dif\lambda \frac{\dif^3 N(\vec p)}{\mathop{\dif\xi} \mathop{\dif\lambda} \mathop{\dif z} } 
    + \sum_i \ln\frac{\frac{\dif^4 N(\boldsymbol p)}{\mathop{\dif\xi} \mathop{\dif\lambda} \mathop{\dif \boldsymbol g_\mathrm{t}} \mathop{\dif z}}
    \Big|_{\xi_i, \lambda_i, \boldsymbol{g}_{\mathrm{t},i}, z_i}}
    {\int_{\lambda_\mathrm{min}(z_i)}^\infty \mathop{\dif\lambda} \frac{\dif^3 N(\boldsymbol p)}{\mathop{\dif\xi} \mathop{\dif\lambda} \mathop{\dif z}}
    \Big|_{\xi_i, z_i}}
    + \mathrm{const.},
\end{equation}
where both sum runs over all clusters in the sample.
The first two terms in Eq.~(\ref{eq:likelihood}) are the Poisson likelihood in $(\xi, z)$-space with the condition $\lambda>\lambda_\mathrm{min}(z)$.
The last term in Eq.~(\ref{eq:likelihood}) is the conditional probability,
\begin{equation}
\label{eq:masscalib}
    \frac{\frac{\dif^4 N(\boldsymbol p)}{\mathop{\dif\xi} \mathop{\dif\lambda} \mathop{\dif \boldsymbol g_\mathrm{t}} \mathop{\dif z}}
    }{\int_{\lambda_\mathrm{min}(z)}^\infty \mathop{\dif\lambda} \frac{\dif^3 N(\boldsymbol p)}{\mathop{\dif\xi} \mathop{\dif\lambda} \mathop{\dif z}}}
    =  \frac{P(\lambda,\, \vec g_\mathrm{t}, \xi, z | \vec p)}{P(\lambda>\lambda_\mathrm{min}(z), \xi, z | \vec p)}
    \equiv P(\lambda,\, \vec g_\mathrm{t} | \lambda>\lambda_\mathrm{min}(z), \xi, z, \vec p),
\end{equation}
which we refer to as the ``mass calibration likelihood''.
Finally, the ``lensing likelihood'' $P(\vec g_\mathrm{t}|M_\mathrm{WL}, \vec p)$ for each cluster is computed as a product of independent Gaussian probabilities in each radial bin $i$
\begin{equation}
    P(\vec g_\mathrm{t}|M_\mathrm{WL}, \vec p) = \prod_i \left(\sigma_{g_{\mathrm{t},i}}\sqrt{2\pi} \right)^{-1}\exp \left[ -\frac12 \left(\frac{g_{\mathrm{t},i} - g_{\mathrm{t},i}(M_\mathrm{WL}, \vec p)}{\sigma_{g_{\mathrm{t},i}}}\right)^2 \right],
\end{equation}
with shape noise $\sigma_{g_{\mathrm{t},i}}$. Note that the model shear profile $\vec g_\mathrm{t}(M_\mathrm{WL}, \vec p)$ explicitly depends on the cosmological parameters in $\vec p$ through the distances in $\Sigma_\mathrm{crit}^{-1}$ [Eq.~(\ref{eq:Sigmacrit})].

\subsection{Numerical Implementation}

We compute
\begin{equation}
  \int_{\lambda_\mathrm{min}(z)}^\infty \mathop{\dif \lambda }\frac{\dif^3 N(\vec p)}{\mathop{\dif\xi} \mathop{\dif\lambda} \mathop{\dif z}} \Big|_{\xi, z} =
  \int_{\lambda_\mathrm{min}(z)}^\infty \mathop{\dif \lambda }
  \iiint \mathop{\dif\Omega_\mathrm{s}} \mathop{\dif M} \mathop{\dif\zeta} \mathop{\dif \tilde\lambda}
  P(\xi|\zeta) P(\lambda|\tilde\lambda)
  P(\zeta, \tilde\lambda |M,z,\vec p)
  \frac{\dif^2 N(M, z, \boldsymbol p)}{\mathop{\dif M} \mathop{\dif V}}
  \frac{\dif^2 V(z,\vec p)}{\mathop{\dif z} \mathop{\dif\Omega_\mathrm{s}}}
\end{equation}
on a regular grid in $(\xi, z)$.
In practice, since $P(\zeta|M,z,\boldsymbol{p})$ is different for each SPT field [Eqs.~(\ref{eq:zetaM}) and~(\ref{eq:aszfield})], we compute a different grid for each SPT field.
Each individual field is homogeneous to good approximation, and the integral $\dif\Omega_\mathrm{s}$ over the field's footprint is separable; we simply multiply with the field area $\Omega_\mathrm{s}$.
For the SPT fields that do not overlap with the DES footprint, the calculation is simpler because it does not involve richness.
With these grids, we can evaluate the first two terms of the log-likelihood given by Eq.~(\ref{eq:likelihood}).
\end{widetext}

The main computational challenge for the analysis pipeline is the evaluation of the mass calibration likelihood Eq.~(\ref{eq:masscalib}) for each cluster in the sample.
While the denominator does not need to be explicitly computed because it can straightforwardly be evaluated from the grid in $(\xi,z)$ we just discussed, the numerator involves the four-dimensional convolution in Eq.~(\ref{eq:d4N_dobs}).
We address the computational challenge with Monte Carlo integration.

The key to efficient Monte Carlo integration is a good sampling of the integration parameter space, meaning that no computation time should be wasted on parts of the integrand that contribute negligibly to the integral.
In a previous SPT analysis, an efficient Monte Carlo integration scheme for the case of X-ray follow-up data $Y_\mathrm{X}$ was presented \cite{dehaan16}.
That algorithm draws random deviates $\zeta$ and $Y_\mathrm{X}$ from the observed quantities $\xi$ and $Y_\mathrm{X}^\mathrm{observed}$, and then draws random deviates for the halo mass from the distribution $P(M|\zeta, Y_\mathrm{X})$.
The value of the integral is then proportional to the mean of the probabilities $P(M) \equiv \frac{\dif^3 N(\boldsymbol p)}{\mathop{\dif M} \mathop{\dif z} \mathop{\dif V}}$ of each random draw.
While this algorithm could be readily applied to the richness follow-up data, it cannot be applied to the lensing data, because $P(M_\mathrm{WL}|\boldsymbol g_\mathrm{t})$ is not properly normalized.
In other words, we cannot in general draw random deviates \Mwl\ given a measured shear profile $\boldsymbol g_\mathrm{t}$.
Consider for example a cluster that has negative shear due to the rather large shape noise; this cannot lead to the random draw of a well defined (i.e., positive) lensing mass \Mwl.
We thus design a new Monte Carlo integration scheme.

Our Monte Carlo integration of Eq.~(\ref{eq:d4N_dobs}) is iterative.
In a first pass, we draw a modest number of $2^{11}=2048$ log-masses uniformly in the wide mass range $10^{13}<M/(h^{-1}\Msun)<10^{16}$.
For each log-mass draw, we draw random deviates $\zeta,\tilde\lambda,M_\mathrm{WL}$ according to the multi-variate scaling relation [Eq.~(\ref{eq:scaling_relation})].
We then evaluate $P(\xi|\zeta)$, $P(\lambda|\tilde\lambda)$, and the lensing likelihood $P(\boldsymbol g_\mathrm{t}|M_\mathrm{WL}, \vec p)$ using the observed quantities $\xi, \lambda, \boldsymbol{g_\mathrm{t}}$.
The (un-normalized) probability of each mass draw $i$ then is
\begin{equation}
  \label{eq:MC_PlnM}
    P(\ln M_i) = P(\xi|\zeta)\,P(\lambda|\tilde\lambda)\, P(\boldsymbol g_\mathrm{t}|M_\mathrm{WL}, \vec p) \, \frac{\dif N(\boldsymbol p)}{\dif \ln M}.
\end{equation}
In a second pass, we now draw a large number of $2^{15}=32,768$ log-masses from the distribution $P(\ln M)$.
By construction, $P(\ln M)$ describes the part of the integrand that has high probability and we thus have constructed an efficient Monte Carlo integrator.
We evaluate the individual contributions as described above, and obtain a final estimate of the integral as
\begin{equation}
    \frac{\dif^4 N(\vec p)}{\mathop{\dif\xi} \mathop{\dif\lambda} \mathop{\dif \vec g_\mathrm{t}} \mathop{\dif z} } = \left\langle \frac{P(\ln M_i)}{\mathrm{prior(\ln M_i)}}\right\rangle,
\end{equation}
where the prior distribution is the distribution the log-masses were drawn from [that is, Eq.~(\ref{eq:MC_PlnM})].

For clusters without weak-lensing measurements we only need to evaluate $P(\lambda| \lambda>\lambda_\mathrm{min}(z), \xi, z, \vec p)$.
The integral then reduces to a lower-dimensional one, and can be solved in an analogous way.
Obviously, for cluster with no lensing or richness data, the term $P(\lambda,\, \vec g_\mathrm{t} | \lambda>\lambda_\mathrm{min}(z), \xi, z, \vec p)$ is constant and does not need to be computed at all.

We note that because the error model for the observed richness is lognormal [see Eq.~(\ref{eq:P_lambda_tildelambda})], the convolution with the observational error does not need to be explicitly computed.
Instead, the observational scatter can be straightforwardly combined with the intrinsic scatter.
In our discussion, we explicitly track $P(\lambda|\tilde\lambda)$ for the purpose of completeness.

\section{Pipeline validation using Mock Catalogs}
\label{sec:validation}

We implement the analysis framework described in this paper as a Python module for \textsc{CosmoSIS} \citep{zuntz15}.\footnote{\url{https://cosmosis.readthedocs.io/}}
We test the pipeline using full-scale mock catalogs that are drawn from the model, verifying that we can recover the input parameters.
The mock catalogs are created by drawing halos from the halo mass function (using Poisson statistics), drawing realizations of the multi-observable scaling relations, and applying the survey selection cuts. The mocks are formatted identically to the real data.
Our validation approach is a meaningful test of the analysis pipeline because creating the mocks is significantly less challenging than implementing the likelihood function.

\begin{figure*}
  \includegraphics[width=\textwidth]{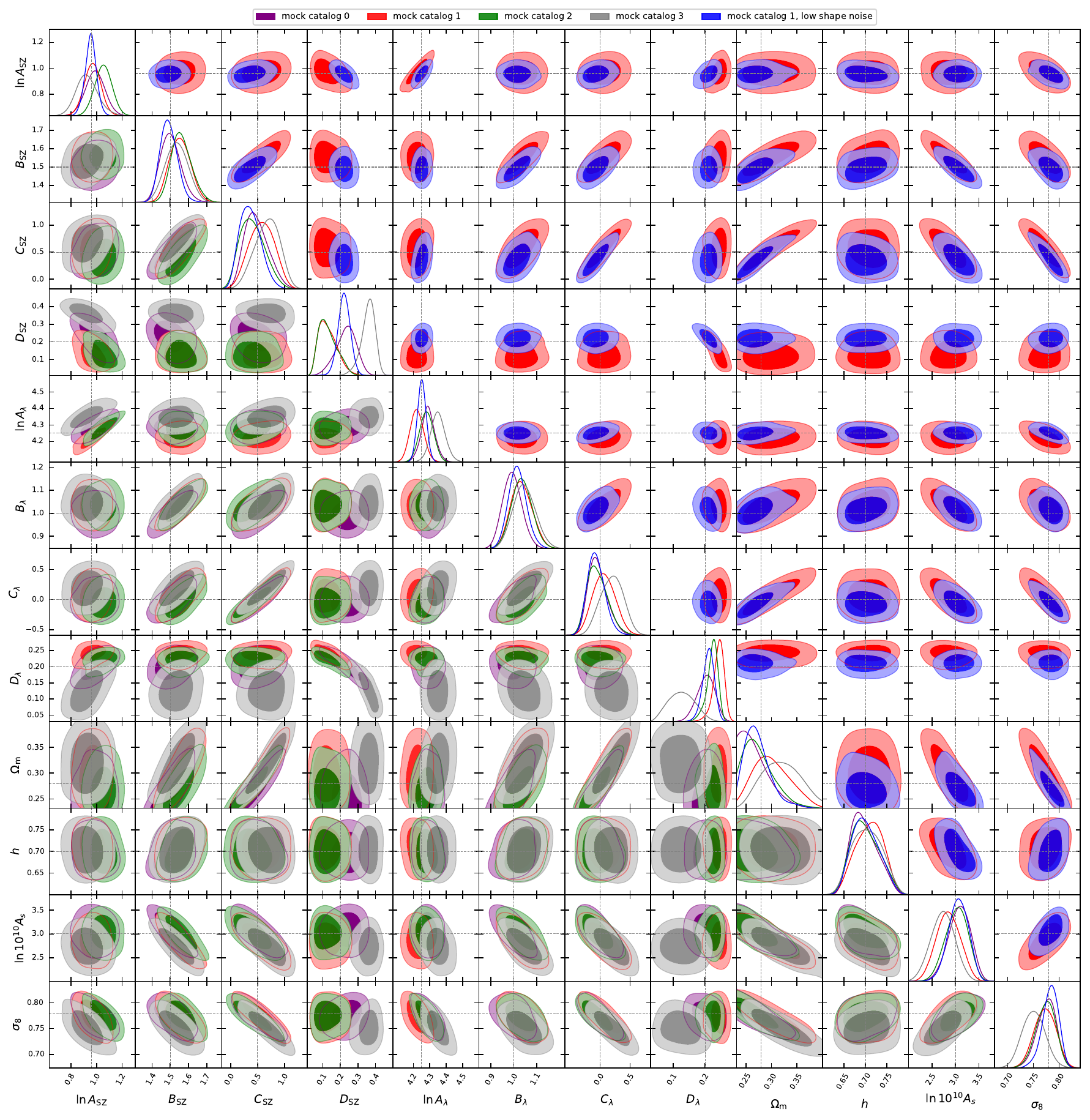}
  \caption{Analysis of four statistically independent mock catalogs in a flat \LCDM\ cosmology.
  There is visible statistical scatter between the different mock catalog realizations.
  Dashed lines show the parameter input values, which are recovered within the uncertainties.
  We apply an informative prior $\mathcal N(0.7, 0.05^2)$ to the Hubble parameter $h$ and require $\Om>0.232$, $\sigma_{\ln\zeta}>0.05$, and $\sigma_{\ln\tilde\lambda}>0.05$.
  All other parameters are marginalized over wide flat ranges.
  Blue contours show the analysis of mock~1, but with mock lensing data that has four times lower shape noise than the fiducial mock (red contours).}
  \label{fig:mock_test}
\end{figure*}

We create four statistically independent mock catalogs by performing the aforementioned steps for a set of different initial random seeds \{0, 1, 2, 3\} (for the same model and the same input parameters).
For the analysis of the mock catalogs, our prime interest is in assessing whether the pipeline has any remaining biases.
We are not necessarily interested in keeping track of all potential sources of uncertainty (which we will, of course, in the analysis of the real data) and so, for simplicity and to make the mock tests slightly more stringent, we fix the parameters of the \Mwl--\Mhalo\ relations, the correlated scatter parameters $\rho$, and the cosmological parameters $\Omega_\mathrm{b}h^2$, $\Omega_\nu h^2$, and $n_s$ to their input values.
Because the cluster abundance data cannot meaningfully constrain the Hubble parameter $h$, we apply a Gaussian prior $h\sim\mathcal N(0.7, 0.05^2)$, centered on the input value of 0.7.
In terms of the cosmological parameters, we thus sample \Om, $\ln(10^{10}A_s)$, and $h$, and record $\sigma_8$ as a derived parameter.

We show the parameter constraints from the mock analyses in Fig.~\ref{fig:mock_test}.
The results show some amount of statistical scatter from catalog to catalog.
The parameter input values are shown with lines, and are recovered within the uncertainties.

To perform a more stringent test, we re-create the lensing data of mock~1 but assume that shape noise is four times lower than in the real data (and in the fiducial mock catalogs).
As expected, the analysis of that mock dataset (named ``mock catalog 1, low shape noise'' in the figure), produces tighter parameter constraints, that still agree with the input parameters.

The validation tests confirm that our analysis pipeline correctly implements our modeling framework.
The pipeline is thus ready to be used for the analysis of the real dataset.
Note that the pipeline test presented here does not answer the question whether the model we implemented is a good description of the real data.
This question cannot be answered using synthetic data.
What we confirm here is that the code correctly reflects the framework described in this paper, and that it is self-consistent in its ability to recover unbiased measurements of cosmological parameters from mock inputs.
In the analysis of the real data in \citetalias{bocquet24II}, we will perform a series of blind tests to verify that the assumed model is indeed able to describe the real dataset.

\section{Summary}
\label{sec:summary}

In this paper, we present the analysis framework that we will use to extract cosmological information from the abundance of clusters detected in the SPT-SZ and SPTpol surveys with a simultaneous mass calibration using weak-lensing data from DES~Y3 and HST.
The results of the analysis of the real data will be presented in \citetalias{bocquet24II}.

We build a Bayesian population model to describe the cluster abundance assuming Poisson statistics, and we forward-model the cluster selection as cuts in the SPT detection significance $\xi$, cluster redshift $z>0.25$, and a cut in optical richness $\lambda_\mathrm{min}(z)$ for the part of the survey footprint that is shared between SPT and DES.
We perform a simultaneous weak-lensing cluster mass calibration on a cluster-by-cluster basis, i.e., we do not stack the lensing signal for multiple clusters. We account for the intrinsic and observational scatters in all cluster observables and allow the intrinsic scatter to be correlated among the observables.

A key focus of this work is to prepare the DES~Y3 lensing data for cluster mass calibration.
We establish a data-driven model for cluster miscentering and find some tension with current hydrodynamic simulations. Our analysis is thus based on the data-driven miscentering model, and we leave a more detailed comparison with simulations for future work.
We set up a flexible model to describe the impact of cluster member contamination. Contaminants are described by a Gaussian distribution that is offset from the cluster redshift. The width of the Gaussian and the amount of offset are free parameters. The radial trend is described by an NFW profile with free concentration. The amount of contamination is modeled as a power law in richness, and as a flexible function of redshift, to accommodate the non-trivial impact of filter band transitions.
We combine these models with the DES~Y3 source redshift distributions and projected mass maps from hydrodynamic simulations to establish an effective model that creates the link between halo mass and the measured shear profiles.
For the current lensing dataset, we estimate an accuracy in lensing mass that varies between 1\% at $z=0.25$ and 10\% at $z=0.95$.
We add an additional 2\% uncertainty due to uncertainties in the impact of hydrodynamic effects in quadrature, and obtain a final accuracy between 2--10\%.
Note that the first set of numbers can be improved by reducing the systematic uncertainties in the source redshift distribution. We thus expect significant progress with the upcoming data from, e.g., the Euclid\footnote{\url{https://www.euclid-ec.org}} and Vera C. Rubin observatories.\footnote{\url{https://www.rubinobservatory.org}}
The additional 2\% uncertainty, however, reflects our current lack of knowledge of how the halo mass distributions are influenced by hydrodynamic effects, and importantly, this estimate is based on the comparison of only two numerical simulations.
More work, and more comparisons between different hydrodynamic feedback models is needed to better characterize and to reduce this uncertainty.

We introduce the multi-observable likelihood function and discuss its implementation in our analysis pipeline. We validate the pipeline, demonstrating that it is able to produce unbiased constraints by analyzing synthetic mock datasets that are drawn from the model.

The analysis framework presented here enables robust cluster cosmology analyses using samples of about 1,000~clusters. It remains to be shown whether our analysis approach can also be efficiently applied to much larger cluster samples selected in optical data or from upcoming, deep X-ray and SZ surveys (e.g., from eROSITA,\footnote{\url{https://www.mpe.mpg.de/eROSITA}} SPT-3G, Simons Observatory,\footnote{\url{https://simonsobservatory.org/}} or CMB-S4\footnote{\url{https://cmb-s4.org/}}) or whether stacking approaches will prove to be more practical.

\begin{acknowledgments}

This research was supported by the Excellence Cluster ORIGINS, which is funded by the Deutsche Forschungsgemeinschaft (DFG, German Research Foundation) under Germany's Excellence Strategy - EXC-2094-390783311, the MPG Faculty Fellowship program and the Ludwig-Maximilians-Universit\"at M\"unchen. Parts of the MCMC computations have been carried out on the computing facilities of the Computational Center for Particle and Astrophysics (C2PAP). 
The Bonn and Innsbruck authors acknowledge support from the German Federal Ministry for Economic Affairs and Energy (BMWi) provided through DLR under projects 50OR2002 and 50OR2302, from the German Research Foundation (DFG) under grant 415537506, and the Austrian Research Promotion Agency (FFG) and the Federal Ministry of the Republic of Austria for Climate Action, Environment, Mobility, Innovation and Technology (BMK) via grants 899537 and 900565.
The Melbourne authors acknowledge support from the Australian Research Council’s Discovery Projects scheme (No. DP200101068).

The South Pole Telescope program is supported by the National Science Foundation (NSF) through the Grant No. OPP-1852617. Partial support is also provided by the Kavli Institute of Cosmological Physics at the University of Chicago.
PISCO observations were supported by US NSF grant AST-0126090. 
Work at Argonne National Laboratory was supported by the U.S. Department of Energy, Office of High Energy Physics, under Contract No. DE-AC02-06CH11357.

Funding for the DES Projects has been provided by the U.S. Department of Energy, the U.S. National Science Foundation, the Ministry of Science and Education of Spain, 
the Science and Technology Facilities Council of the United Kingdom, the Higher Education Funding Council for England, the National Center for Supercomputing 
Applications at the University of Illinois at Urbana-Champaign, the Kavli Institute of Cosmological Physics at the University of Chicago, 
the Center for Cosmology and Astro-Particle Physics at the Ohio State University,
the Mitchell Institute for Fundamental Physics and Astronomy at Texas A\&M University, Financiadora de Estudos e Projetos, 
Funda{\c c}{\~a}o Carlos Chagas Filho de Amparo {\`a} Pesquisa do Estado do Rio de Janeiro, Conselho Nacional de Desenvolvimento Cient{\'i}fico e Tecnol{\'o}gico and 
the Minist{\'e}rio da Ci{\^e}ncia, Tecnologia e Inova{\c c}{\~a}o, the Deutsche Forschungsgemeinschaft and the Collaborating Institutions in the Dark Energy Survey. 

The Collaborating Institutions are Argonne National Laboratory, the University of California at Santa Cruz, the University of Cambridge, Centro de Investigaciones Energ{\'e}ticas, 
Medioambientales y Tecnol{\'o}gicas-Madrid, the University of Chicago, University College London, the DES-Brazil Consortium, the University of Edinburgh, 
the Eidgen{\"o}ssische Technische Hochschule (ETH) Z{\"u}rich, 
Fermi National Accelerator Laboratory, the University of Illinois at Urbana-Champaign, the Institut de Ci{\`e}ncies de l'Espai (IEEC/CSIC), 
the Institut de F{\'i}sica d'Altes Energies, Lawrence Berkeley National Laboratory, the Ludwig-Maximilians-Universit{\"a}t M{\"u}nchen and the associated Excellence Cluster Origins, 
the University of Michigan, NSF's NOIRLab, the University of Nottingham, The Ohio State University, the University of Pennsylvania, the University of Portsmouth, 
SLAC National Accelerator Laboratory, Stanford University, the University of Sussex, Texas A\&M University, and the OzDES Membership Consortium.

Based in part on observations at Cerro Tololo Inter-American Observatory at NSF's NOIRLab (NOIRLab Prop. ID 2012B-0001; PI: J. Frieman), which is managed by the Association of Universities for Research in Astronomy (AURA) under a cooperative agreement with the National Science Foundation.

The DES data management system is supported by the National Science Foundation under Grant No. AST-1138766 and AST-1536171.
The DES participants from Spanish institutions are partially supported by MICINN under Grants ESP2017-89838, PGC2018-094773, PGC2018-102021, SEV-2016-0588, SEV-2016-0597, and MDM-2015-0509, some of which include ERDF funds from the European Union. IFAE is partially funded by the CERCA program of the Generalitat de Catalunya.
Research leading to these results has received funding from the European Research
Council under the European Union's Seventh Framework Program (FP7/2007-2013) including ERC Grant agreements 240672, 291329, and 306478.
We  acknowledge support from the Brazilian Instituto Nacional de Ci\^encia
e Tecnologia (INCT) do e-Universo (CNPq Grant 465376/2014-2).

This manuscript has been authored by Fermi Research Alliance, LLC under Contract No. DE-AC02-07CH11359 with the U.S. Department of Energy, Office of Science, Office of High Energy Physics.

This research has made use of the SAO/NASA Astrophysics Data System and of adstex.\footnote{\url{https://github.com/yymao/adstex}}
We wish to thank the referee for their thoughtful comments.
We thank Ludwig, the SPT support cat for this analysis. Inquiries about SPT support cats shall be directed to T.C.

\end{acknowledgments}

\appendix
\section{Lensing Efficiencies of the DES~Y3 Source Bins}
\label{sec:lens_eff_ratio}

\begin{figure}
  \includegraphics[width=\columnwidth]{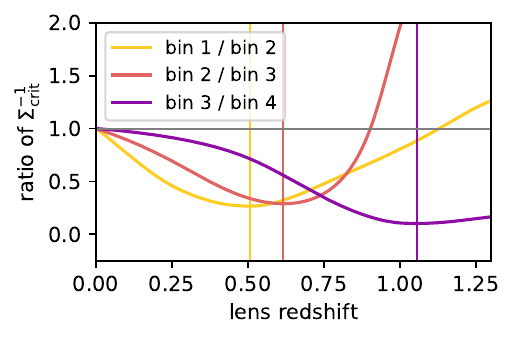}
  \caption{Ratio of lensing efficiencies $\Sigma_\mathrm{crit}^{-1}$ for the tomographic source bins. Vertical lines mark the lens redshift at which the ratio stops decreasing.
  }
  \label{fig:invSigmacrit_ratio}
\end{figure}

In Fig.~\ref{fig:invSigmacrit_ratio}, we show how the ratio of lensing efficiencies $\langle\invSigmac\rangle$ between subsequent DES~Y3 source bins evolve as a function of lens redshift.
The bins were originally chosen to have increasing mean redshifts, and one would thus expect a high-redshift bin to also have a higher lensing efficiency.
Interestingly, however, we observe that at lens redshift $z\sim1.1$, the lensing efficiency of source bin~1 becomes larger than that of bin~2. Similarly, at $z\sim0.9$, the lensing efficiency of source bin~2 becomes larger than that of bin~3.
It does not seem advisable to use the lensing dataset in this regime which clearly does not behave as intended.
A more robust redshift limit for each source bin may be estimated as the redshift at which the ratio of the lensing efficiencies of, e.g., bin~1 and bin~2 starts rising (yellow line at $z\sim0.5$ in Fig.~\ref{fig:invSigmacrit_ratio}, and orange line at $z\sim0.6$ for the ratio of bin~2 and bin~3).

In our analysis, we use a given source bin only for lenses with redshifts that are smaller than the median source redshift of that bin.
This requirement is more stringent than the discussion of ratios of lensing efficiencies presented here, and we conclude that our analysis is robust to the potential problems addressed in this appendix.

Finally, we remind the reader that the source bins were originally defined for the 3x2~pt analysis.
That analysis is not very sensitive to the high-redshift tails of the source redshift distributions but rather to an accurate calibration of the mean redshift.
In this appendix, we thus explore the lensing data products in a regime that was not validated.
For future analyses of lensing datasets from wide-field surveys such as Euclid and LSST, we recommend that multiple use cases including the analysis of galaxy clusters, cosmic shear, galaxy-galaxy lensing, and other lensing probes be considered jointly.

\section{Joint SZ, optical, and X-ray miscentering}
\label{sec:X-ray_miscenter}

\begin{figure*}
  \includegraphics[width=\textwidth]{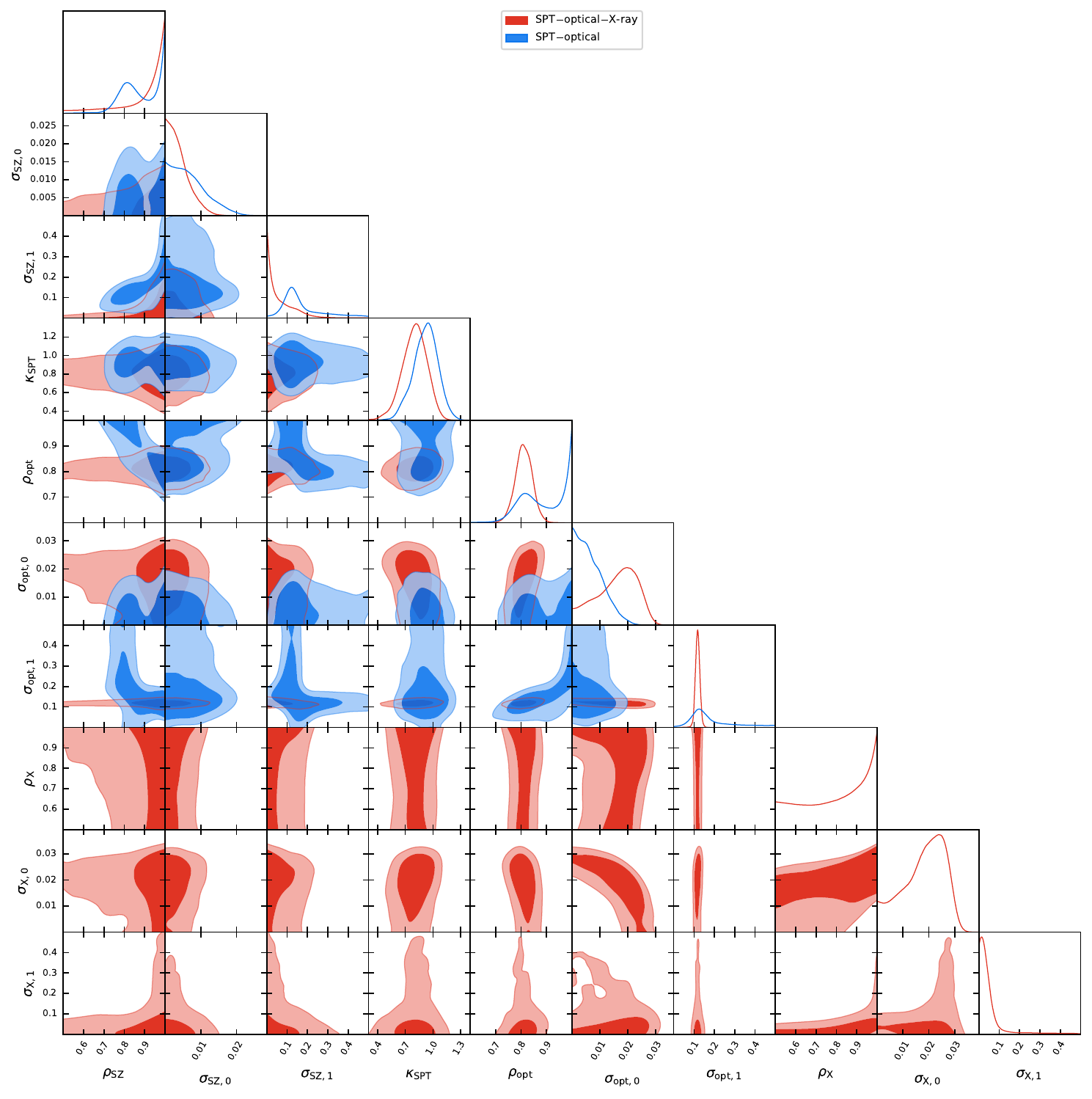}
  \caption{Constraints (68\% and 95\% credible regions) on the parameters of the miscentering model.
  Blue contours are obtained by analyzing the SZ--optical offsets and fitting for the parameters of the SZ--true and optical--true miscentering distributions.
  Red contours are obtained by analyzing the SZ--optical, SZ--X-ray, and X-ray--optical offsets and fitting for the parameters of all three miscentering distributions.
  Adding the X-ray center helps break some of the degeneracies in the SZ--optical fit.
  Note that we assume that all miscentering distributions are independent, which is particularly problematic in the case of SZ and X-ray centers.
  Therefore, our cluster lensing is based on the results of the more conservative SZ--optical analysis.}
  \label{fig:miscenter_GTC}
\end{figure*}

Cluster X-ray centers have often been used as a proxy for the true halo center, because of the excellent angular resolution and the fact that in hydrostatic equilibrium the peak ICM emission occurs at the minimum of the cluster potential.
We expand the analysis presented in Sec.~\ref{sec:SPT-optical_miscenter} by incorporating 70~large-scale X-ray centroid measurements from Chandra data \citep{mcdonald13, mcdonald17}.
We do not assume that those centroids coincide with the true halo centers and describe the intrinsic X-ray--true offset as in Eq.~(\ref{eq:miscenter}).
We expand the likelihood function by also considering the measured offsets between optical and X-ray centers (modeled as the convolution of the optical--true and X-ray--true offset distributions) and SPT and X-ray centers (modeled as the convolution of the SZ--true and X-ray--true offset distributions and the SPT positional uncertainty).

The recovered parameter constraints on the SZ and optical miscentering are consistent with our baseline results, but they are somewhat tighter (see Fig.~\ref{fig:miscenter_GTC} and Table~\ref{tab:X-ray_miscentering}).
In particular, the cross-shaped degeneracy between $\sigma_{\mathrm{SZ},1}$ and $\sigma_{\mathrm{opt},1}$ is broken.
However, we note that this simplified model does not account for the expected correlation between SZ and X-ray centers or the tendency of optically determined center positions to align with the X-ray center.
Our cosmological analysis can be self-consistently performed using the miscentering distributions calibrated without X-ray data, and the contribution of the uncertainty in the offset modeling to the overall error budget in the lensing mass calibration is negligible (Sec.~\ref{sec:massbiasscatterdiscussion}).
Therefore, we leave further explorations of multi-observable cluster miscentering that includes X-ray observations to future work.

\begin{table}
\caption{\label{tab:X-ray_miscentering}
Parameters of the joint SZ--optical--X-ray miscentering distributions (mean and 68\% credible interval, one-sided limits are for the 95\% credible interval).}
\begin{ruledtabular}
\begin{tabular}{ll}
Parameter & Constraint\\
\colrule
$\rho_\mathrm{SZ}$ & $0.90~(>0.50)$\\
$\sigma_{\mathrm{SZ},0}\, [h^{-1}\mathrm{Mpc}]$ & $0.004^{+0.001}_{-0.004}$\\
$\sigma_{\mathrm{SZ},1}\, [h^{-1}\mathrm{Mpc}]$ & $0.065^{+0.014}_{-0.065}$\\
$\kappa_\mathrm{SPT}$ & $0.80^{+0.14}_{-0.12}$\\
\colrule
$\rho_\mathrm{opt}$ & $0.81^{+0.03}_{-0.03}$\\
$\sigma_{\mathrm{opt},0}\, [h^{-1}\mathrm{Mpc}]$ & $0.016^{+0.009}_{-0.005}$\\
$\sigma_{\mathrm{opt},1}\, [h^{-1}\mathrm{Mpc}]$ & $0.118^{+0.011}_{-0.011}$\\
\colrule
$\rho_\mathrm{X}$ & $0.80\,(>0.50)$\\
$\sigma_{\mathrm{X},0}\, [h^{-1}\mathrm{Mpc}]$ & $0.018^{+0.010}_{-0.005}$\\
$\sigma_{\mathrm{X},1}\, [h^{-1}\mathrm{Mpc}]$ & $0.056\,(<0.248)$\\
\end{tabular}
\end{ruledtabular}
\end{table}

\section{Impact of Sample Variance}
\label{sec:samplevariance}

In the SPT analyses to date, the effect of sample variance has been negligible compared to the more important shot noise (Poisson error).
Since we are now using significantly deeper data (over the SPTpol~500d footprint), we re-assess the situation.

\begin{figure}
  \includegraphics[width=\columnwidth]{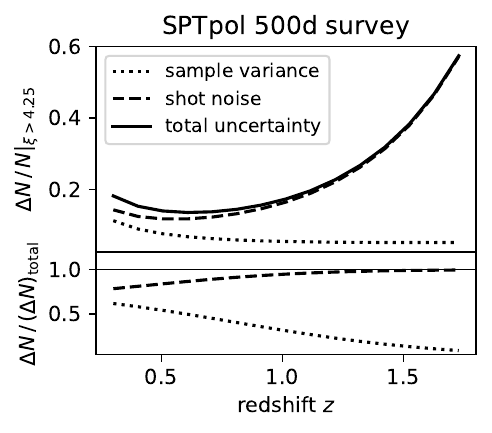}
  \caption{Contributions of sample variance and shot noise to the total uncertainty in the cluster abundance in our deepest SPT field.}
  \label{fig:sample_var}
\end{figure}

We compute the sample variance in the predicted cluster abundance in SPTpol~500d, the deepest patch of our survey (following, e.g., \cite{hu&kravtsov03, valageas11}).
For simplicity, we assume that the field's footprint is circular on the sky.
We assume a fiducial cosmology and scaling relation parameters and apply the cluster selection with $z>0.25$ and $\xi>4.25$.
In Fig.~\ref{fig:sample_var}, we show the contributions to the relative uncertainty in the predicted cluster abundance due to shot noise and the sample variance.
For all redshifts, the contribution from sample variance is smaller than shot noise.
All other SPT fields are significantly shallower, and the relative importance of sample variance is even smaller.
We thus neglect the effect of sample variance on the SPT cluster abundance.

\bibliographystyle{apsrev}
\bibliography{paperI.bib}

\end{document}